\documentclass{aa}  

\usepackage{graphicx}
\usepackage{subcaption}
\usepackage{txfonts}
\usepackage{xcolor}
\usepackage{bbold}
\usepackage[extra]{tipa}
\usepackage[normalem]{ulem}

\usepackage{hyperref}

\usepackage{siunitx}
\DeclareSIUnit \jy {Jy}

\begin{document} 
    \titlerunning{Impact of diffuse Galactic emission on direction-independent gain calibration}
    \authorrunning{H{\"o}fer et al.}
   \title{The impact of diffuse Galactic emission on direction-independent gain calibration in high-redshift 21 cm observations}

   \author{C. H{\"o}fer
          \inst{1}\fnmsep\thanks{e-mail: hofer@astro.rug.nl},
          L.V.E. Koopmans
          \inst{1},
          S.A. Brackenhoff
          \inst{1},
          E. Ceccotti
          \inst{1,2},
          K. Chege
          \inst{1},
          S. Ghosh
          \inst{1},
          F.G. Mertens
          \inst{1,3},
          M. Mevius
          \inst{4},
          S. Munshi
          \inst{1}
          \and
          A.R. Offringa
          \inst{4}
          }
    \institute{Kapteyn Astronomical Institute, University of Groningen, PO Box 800, 9700 AV Groningen, The Netherlands
    \and INAF -- Instituto di Radioastronomia, Via P. Gobetti 101, 40129 Bologna, Italy
    \and LUX, Observatoire de Paris, PSL Research University, CNRS, Sorbonne Universit{\'e}, F-75014 Paris, France
    \and Astron, PO Box 2, 7990 AA Dwingeloo, The Netherlands}

   \date{Received XX March, 2025; accepted XX XX, 2025}

 
  \abstract
   {This study examines the impact of diffuse Galactic emission on sky-based direction-independent (DI) gain calibration using realistic forward simulations of Low-Frequency Array (LOFAR) observations of the high-redshift \SI{21}{\centi\metre} signal of neutral hydrogen during the Epoch of Reionization (EoR). We simulated LOFAR observations between \SIrange{147}{159}{\mega\hertz} using a sky model that includes a point source catalog and diffuse Galactic emission. The simulated observations were DI-gain calibrated with the point source catalog alone, utilizing the LOFAR-EoR data analysis pipeline. A full power spectrum analysis was conducted to measure the systematic bias, relative to thermal noise, caused by DI-gain calibration using a point-source-only (PSO) sky model, when applied to simulated data that include both point sources and diffuse Galactic emission. The results are compared to a ground truth scenario where both the simulated sky and the calibration model include only point sources. Additionally, the cross-coherence between observation pairs was computed to determine whether DI-gain calibration errors are coherent or incoherent in specific regions of power spectrum space as a function of integration time. We find that DI-gain calibration with a PSO sky model that omits diffuse Galactic emission introduces a systematic bias in the power spectrum for $k_{\parallel}$ bins < 0.2 $h\,\mathrm{Mpc}^{-1}$. The power spectrum errors in these bins are coherent in time and frequency; therefore, the resulting bias could be mitigated during the foreground removal step using Gaussian Process Regression, as demonstrated in previous studies. In contrast, errors for $k_{\parallel}$ > 0.2 $h\,\mathrm{Mpc}^{-1}$ are largely incoherent and average down as noise. We conclude that, based on our analysis prior to foreground removal, missing diffuse Galactic emission in the sky model during DI-gain calibration is unlikely to be a dominant contributor to the excess noise observed in the current LOFAR-EoR upper limits on the 21 cm signal power spectrum.}

   \maketitle
%

\section{Introduction}
\label{sec:intro}
One of the key questions from the early universe is how and when the first luminous sources – such as stars, galaxies, and quasars – formed and how their radiation ionized the surrounding intergalactic medium. The Cosmic Dawn (CD, $z \sim 15 - 30$) and the Epoch of Reionization (EoR, $z\sim 6 -15$) refer to the eras when the first stars and galaxies formed and began emitting enough ultraviolet radiation, ionizing neutral hydrogen in the intergalactic medium (IGM) and transforming it from a neutral to a fully ionized state. Measurements of the optical depth in Cosmic Microwave Background (CMB) observations \citep{2013:hinshaw, 2020:Planck}, the Gunn-Peterson trough observed in quasar absorption spectra \citep{2001:becker} and Lyman-alpha emission from high-redshift galaxies \citep{2010:Ouchi, 2010:stark} support the picture that reionization began around $z \sim 10$ and was largely complete by $z \sim 6$. 

Observations of the CD and EoR are still limited in number, primarily capturing information from the brightest objects observed by telescopes in the optical and infrared wavelengths. The James Webb Space Telescope (JWST) is pushing these boundaries and observing galaxies at redshifts as high as $z \sim 15$ \citep{2022:atek, 2022:donnan, 2023:harikane, 2024:finkelstein}. The detection of such bright and early galaxies suggests that star formation occurred in more massive or more intensely star-forming regions, and much earlier than previously expected. 

The most comprehensive probe of the CD and EoR is the measurement of the redshifted \SI{21}{\centi\metre} line of neutral hydrogen, either in emission or absorption against the CMB. This is done as a function of frequency and angular scale using radio telescopes to map the 3D distribution of matter in the universe. Many experiments are currently underway, either measuring the sky-averaged \SI{21}{\centi\metre} brightness temperature with single receiver instruments such as EDGES\footnote{Experiment to Detect the Global Epoch of Reionization Signature, \href{https://loco.lab.asu.edu/edges}{https://loco.lab.asu.edu/edges}} \citep{2018:bowman} and SARAS\footnote{Shaped Antenna measurement of the background Radio Spectrum, \href{http://www.rri.res.in/DISTORTION/saras.html}{http://www.rri.res.in/DISTORTION/saras.html}} \citep{2018:singh}, or the spatially varying \SI{21}{\centi\metre} brightness temperature fluctuations with radio interferometers such as LOFAR\footnote{LOw-Frequency ARray, \href{http://www.lofar.org}{http://www.lofar.org}} \citep{2013:van_Haarlem}, MWA\footnote{Murchison Widefield Array, \href{http://www.mwatelescope.org}{http://www.mwatelescope.org}} \citep{2013:tingay}, HERA\footnote{Hydrogen Epoch of Reionization Array, \href{http://reionization.org}{http://reionization.org}} \citep{2017:deboer}, NenuFAR\footnote{New Extension in Nançay
Upgrading LOFAR, \href{https://nenufar.obs-nancay.fr}{https://nenufar.obs-nancay.fr}} \citep{2020:zarka} and GMRT\footnote{Giant Metrewave Radio Telescope, \href{http://gmrt.ncra.tifr.res.in}{http://gmrt.ncra.tifr.res.in}} \citep{2017:gupta}. The latter instruments have set competitive upper limits on the \SI{21}{\centi\metre} signal power spectrum, yet a detection is still missing. The most stringent published upper limits on the \SI{21}{\centi\metre} spectrum during the EoR are as follows. The MWA reported a 2-$\sigma$ upper limit of $\Delta^2_{21} < (\SI{43.1}{\milli\kelvin})^2$ at $k$ = 0.14 $h\,\mathrm{Mpc}^{-1}$ and $z \approx 6.5$ \citep{2020:trott}, HERA reported a 2-$\sigma$ upper limit of $\Delta^2_{21} < (\SI{21.4}{\milli\kelvin})^2$ at $k$ = 0.34 $h\,\mathrm{Mpc}^{-1}$ and $z \approx 7.9$ \citep{2023:hera} and the LOFAR-EoR Key Science Project set a 2-$\sigma$ upper limit of $\Delta^2_{21} < (\SI{72.9}{\milli\kelvin})^2$ at $z \approx 9.1$ \
\citep{2020:mertens}. More recently, updated results from the LOFAR-EoR project have demonstrated a two- to four-fold improvement over previous LOFAR-EoR limits \citep{2025:mertens}. The most stringent 2$\sigma$ upper limits achieved are $\Delta^2_{21} < (\SI{68.7}{\milli\kelvin})^2$ at $k$ = 0.076 $h\,\mathrm{Mpc}^{-1}$ and $z \approx 10.1$, $\Delta^2_{21} < (\SI{54.3}{\milli\kelvin})^2$ at $k$ = 0.076 $h\,\mathrm{Mpc}^{-1}$ and $z \approx 9.1$, and $\Delta^2_{21} < (\SI{65.5}{\milli\kelvin})^2$ at $k$ = 0.083 $h\,\mathrm{Mpc}^{-1}$ and $z \approx 8.3$.

Astrophysical foregrounds from our own Galaxy and extragalactic radio sources, which are orders of magnitude brighter than the faint \SI{21}{\centi\metre} signal, make detection extremely challenging. Moreover, the chromaticity of radio interferometers complicates component separation by mixing angular into frequency structures, an effect known as mode-mixing \citep{2012:morales}, requiring impeccable calibration precision of the frequency-dependent response of the interferometer and removal of foregrounds. The effects of an incomplete sky model during calibration \citep{2016:patil, 2017:ewallwice, 2016:barry}, band-pass calibration and cable reflections \citep{2016:beardsley}, polarization leakage into Stokes-I through Faraday rotation due to imperfect calibration of the instrumental polarization response \citep{2010:jelic, 2018:spinelli}, ionospheric disturbances \citep{2010:koopmans, 2016:vedantham, 2017:jordan, 2024:Brackenhoff}, gridding artifacts during imaging \citep{2019b:offringa}, multi-path propagation \citep{2020:Kern}, and radio frequency interference (RFI) \citep{2019a:offringa, 2019:wilensky} must be meticulously understood in order to detect the redshifted \SI{21}{\centi\metre} signal.

\citet{2016:patil}, \citet{2017:ewallwice} and \citet{2016:barry} conducted detailed studies on the impact of incomplete sky models on calibration, particularly the effect of faint sources below the confusion limit, showing that unmodeled sources introduce significant chromatic calibration errors that contaminate the \SI{21}{\centi\metre} power spectrum across all baselines. Furthermore, large-scale diffuse emission – prominent on baselines shorter than 250 $\lambda$ at frequencies of $\sim$\SI{150}{\mega\hertz} and overlapping with the scales of the \SI{21}{\centi\metre} signal – can be absorbed into the gain solutions during direction-dependent (DD) calibration. This absorption occurs due to overfitting when using all baselines \citep{2019:Sardarabadi, 2020:mertens, 2022:mevius}.

To address these issues, the LOFAR-EoR Key Science Project (KSP) team has implemented a refined calibration approach for the North Celestial Pole (NCP) deep field that:
\begin{enumerate}
\item Applies a baseline cut, using only baselines between \mbox{$250 - 5000\,\lambda$} during station-based DD calibration, while the \SI{21}{\centi\metre} power spectrum is computed from baselines between $50 - 250\,\lambda$.
\item Regularizes gain solutions to maintain a smooth frequency response, minimizing signal suppression and reducing excess noise.
\end{enumerate}
However, excluding shorter baselines from calibration increases the excess noise power on these baselines despite regularization, known as the bias-variance trade-off \citep{2016:patil, 2019:Sardarabadi, 2022:mevius}. While this is a drawback, the key advantage is that the \SI{21}{\centi\metre} signal is not absorbed into the gain solutions, preventing signal suppression.

Despite significant progress in LOFAR-EoR's data analysis strategies and the publication of scientifically relevant upper limits, the measurements still exceed the expected thermal noise power, resulting in excess variance \citep{2017:patil}. The LOFAR-EoR team has intensively investigated the origin of this excess variance over the past years. Studies have examined residual foreground emission from off-center sources \citep{2022:gan}, chromatic direction-independent and direction-dependent gain calibration errors, low-level RFI \citep{2020:mertens}, improved spectral index modeling of extended sources \citep{2023:ceccotti} and ionospheric disturbances \citep{2024:Brackenhoff}, but no definitive cause has been clearly identified.

In this paper, we focus specifically on the impact of unmodeled, unpolarized diffuse Galactic emission on direction-independent (DI) gain calibration of LOFAR-EoR data by means of end-to-end simulations. We adopt the strategy of \citet{2020:mertens}, in which DI-gain calibration uses all baselines between $50 - 5000~\lambda$. Diffuse Galactic emission remains prominent on baselines up to $250~\lambda$, potentially introducing calibration errors on fine frequency scales. Since DI-gain solutions are applied directly to the visibilities without subtracting a sky model, any chromatic calibration errors arising from sky model incompleteness cannot be mitigated in subsequent data analysis steps. Such a study has not been performed previously but is crucial for assessing the impact of DI-gain calibration errors on the power spectrum.

The paper is organized as follows. Section \ref{sec:instrument} provides a brief overview of the LOFAR instrument in the context of EoR observations. In Section \ref{sec:methods}, we introduce the design of our realistic simulation pipeline, \texttt{Simple}, and outline the main steps of the LOFAR-EoR data analysis pipeline. Section \ref{sec:calibration_errors} details the description and analysis of DI-gain calibration errors due to unmodeled diffuse emission. In Section \ref{sec:results}, we demonstrate the impact of DI-gain calibration errors from unmodeled diffuse Galactic emission on power spectra. Finally, Section \ref{sec:summary} presents our summary and conclusions.

\section{Instrument and Observations}
In this section, we briefly introduce the instrument and the observations, which provide the framework and specifications for the end-to-end simulations presented in this paper.
\label{sec:instrument}
\subsection{Instrument}
The Low-Frequency Array \citep{2013:van_Haarlem} is a radio interferometer with its core located in Exloo,
the Netherlands, operating at observing frequencies \SIrange{10}{240}{\mega\hertz}. Observations of the EoR are done using receivers in the High Band Antenna (HBA) 
system which was designed for the  \SIrange{110}{250}{\mega\hertz} frequency range. The HBA system combines 16 dual-polarization antennae together into a square 
\qtyproduct{5 x 5}{\m} "tile" with built-in amplifiers and an analog beam-former, which forms a "tile  beam" with a large field of view (FoV). Tiles are closely packed in 
groups of 24, 48 or 96 into a "station" and classified as either core, remote or international, respectively, depending on their distance from the center of the array. 
The 24 core stations are distributed over an area of \qtyproduct{2x2}{\kilo\metre}, while the 14 remote stations are spread over an area of about \SI{40}{\kilo\metre} East-West 
and \SI{70}{\kilo\metre} North-South. At station level, the signals from 
individual tiles are combined digitally into a phased array using beamforming techniques, which allows digital pointing and tracking of the radio telescope. For remote stations only the inner 24 tiles are 
used in the beam-former in order to give both core and remote stations similar primary beams. The "station beam" has a field of view of $\sim$ \SI{4.1}{^\circ} at \SI{150}{\mega\hertz}. 
A fiber network brings the signals from the stations to the correlator, located at the computing center of the University of Groningen. The data are channelized to a frequency resolution of \SI{3.1}{\kilo\hertz} (64 channels per sub-band), integrated to a time resolution of \SI{2}{\s}, and then multiplied to form visibilities.

\subsection{Observational data}
The LOFAR-EoR KSP observes mainly two deep fields: the North Celestial Pole (NCP) and the field surrounding the bright compact radio source 3C196. 
Observations of the NCP are advantageous for two specific reasons: firstly, the phase and pointing center at DEC = \SI{90.0}{^\circ} is a fixed 
location in the sky, hence tracking of the field is not necessary. Secondly, the NCP can be observed every night throughout the year maximizing observation time.
A total of $\sim$ 2480 hours have been observed with the LOFAR-HBA system for the EoR KSP using all 48 core stations and all 14 remote stations in the frequency 
range \SIrange{115}{189}{\mega\hertz}.  Upper limits on the \SI{21}{\centi\metre} signal have been published from LOFAR observations of the North Celestial Pole (NCP), including 13 hours of data in the redshift range $z \sim$ 9.6 -- 10.6 \citep{2017:patil}, 141 hours in the range $z \sim$ 8.7 -- 9.6 \citep{2020:mertens}, and 200 hours covering $z \sim$ 8.3 -- 10.1 \citep{2025:mertens}. A significant fraction of the available data, however, remains unprocessed. Given the measured excess noise in the power spectra, as mentioned in Section \ref{sec:intro}, the focus was on investigating the source of the excess noise and improving data processing techniques, rather than analyzing additional data. More data will be processed in the coming year. 

\section{Methods}
\label{sec:methods}
We built a realistic forward simulation pipeline named \texttt{Simple}\footnote{\url{https://github.com/cahofer/simple}} (short for: SIMulation Pipeline for LOFAR-EoR) based on LOFAR-EoR NCP observations with 
\texttt{Nextflow}\footnote{\url{https://www.nextflow.io}}. \texttt{Nextflow} is a workflow management tool designed to create, manage, and execute complex computational pipelines. \texttt{Nextflow} 
uses a domain-specific language (DSL) that allows users to define individual tasks and connect them within a pipeline structure. Since the LOFAR-EoR project 
combines software packages written in various programming languages (e.g., Python, C++), \texttt{Nextflow} is particularly beneficial when managing these workflows 
offering flexibility with task isolation, dependency management, and environment configuration. The current LOFAR-EoR analysis pipeline \texttt{NextLeap} was developed using \texttt{Nextflow}, making the interplay between simulations and their analysis particularly user-friendly. The main tasks 
of the simulation pipeline include: 
\begin{enumerate}
    \item Prediction of visibilities from a point-source sky model using the software \texttt{SAGECAL-CO}\footnote{\url{https://github.com/nlesc-dirac/sagecal}} \citep{2015:yatawatta}
    \item Prediction of visibilities from the diffuse Galactic emission using the software \texttt{WSCLEAN}\footnote{\url{https://gitlab.com/aroffringa/wsclean}} \citep{2014:offringa}
    \item Prediction of thermal noise using the Python package \texttt{losito}\footnote{The LOFAR simulation tool \url
{https://github.com/darafferty/losito}} \citep{2021:edler}
\end{enumerate}
The LOFAR-EoR KSP data processing pipeline, implemented in \texttt{NextLeap}, normally includes the following steps, (the software names used in brackets): 
\begin{enumerate}
    \item Pre-processing and RFI excision (\texttt{AOFlagger}\footnote{\url{https://gitlab.com/aroffringa/aoflagger}} \citep{2012:offringa})
    \item Data averaging (Default pre-processing pipeline – \texttt{DP3}\footnote{\url{https://github.com/lofar-astron/DP3}} \citep{2018:vandiepen})
    \item Direction-independent (DI) gain calibration (\texttt{SAGECAL-CO})
    \item  Direction-dependent (DD) gain calibration including sky model subtraction (\texttt{SAGECAL-CO})
    \item Imaging (\texttt{WSCLEAN})
    \item Gridding of visibilities (\texttt{pspipe}\footnote{\url{https://gitlab.com/flomertens/pspipe}})
    \item Power spectrum estimation (\texttt{pspipe})
    \item Removal of residual foregrounds with Gaussian process regression (\texttt{pspipe})
\end{enumerate}

To analyze forward-simulated data and study DI-gain calibration errors caused by missing diffuse emission, steps 1, 4 and 8 can be omitted from our analysis pipeline. This is because we compare our results to a ground truth power spectrum derived from visibilities that do not include DI-gain calibration errors caused by missing diffuse emission in the calibration model. Also, data averaging is performed prior to predicting visibilities from a sky model to reduce computation time.

\subsection{Simulation pipeline}
This section describes the simulation pipeline used to study LOFAR-EoR observations, which includes creating a simplified NCP sky model with 684 components and a diffuse Galactic emission model. The pipeline predicts visibilities of the point-source sky with \texttt{SAGECAL-CO}, diffuse emission with \texttt{WSCLEAN}, and thermal noise using the \texttt{losito} package.
\subsubsection{Point source sky model}
\begin{figure}
    \includegraphics[trim={9cm 0 1cm 0},clip, scale=0.35]{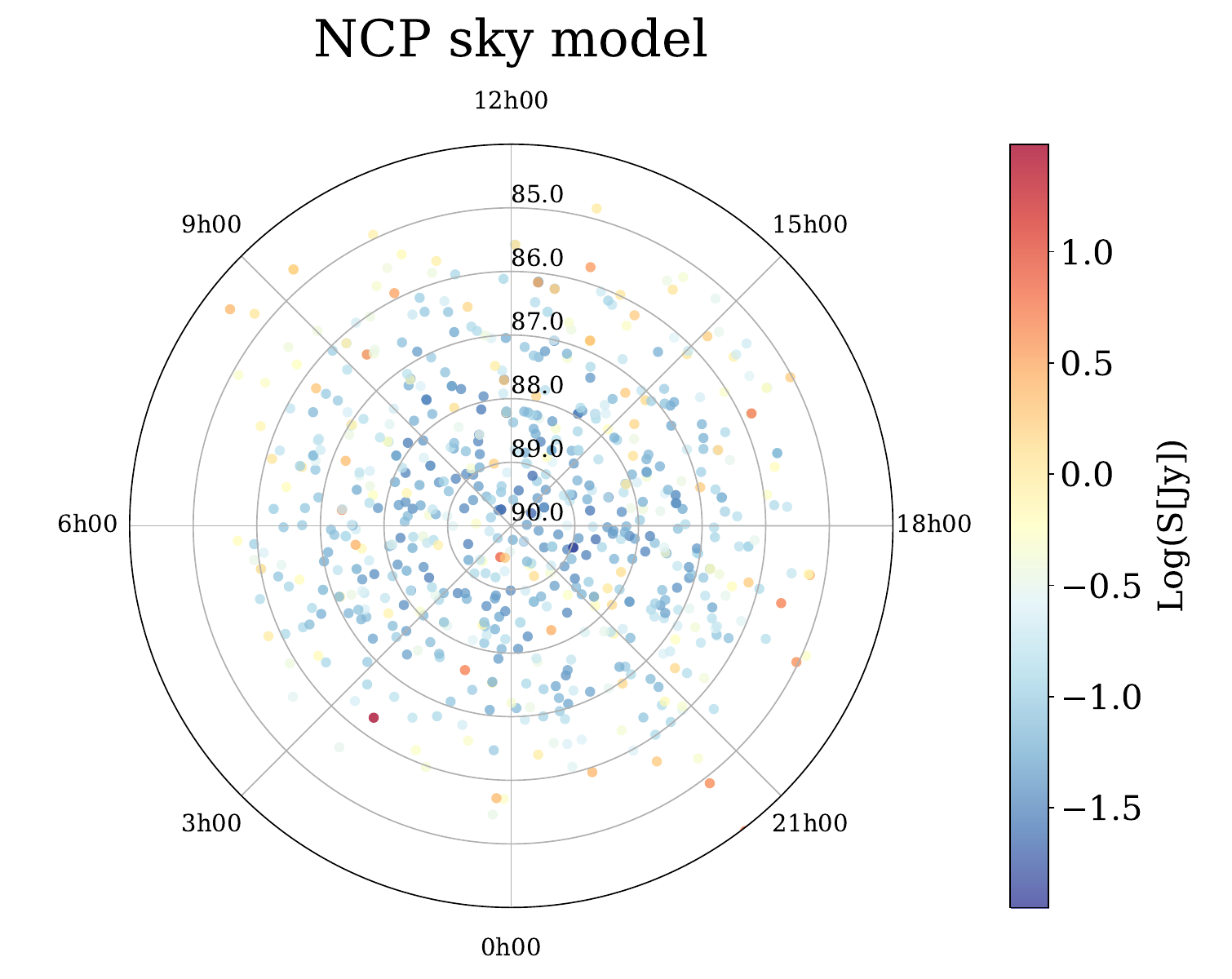}
    \caption{The sky model of the North Celestial Pole at \SI{141}{\mega\hertz} input to the forward simulations. The flux densities are intrinsic and beam corrected. It is composed of 684 unpolarized point sources with a flat spectrum. The bright radio source at DEC $\sim$ \SI{86.3}{^\circ} and RA $\sim$ \SI{2.5}{\hour} is 3C 61.1.}
    \label{fig:skymodel}
\end{figure}

The NCP sky model used for calibrating real LOFAR-EoR observations is a well-developed model composed of 28778 unpolarized components extending \SI{19}{^\circ} 
from the NCP down to an apparent flux density of $\sim$ \SI{3}{\milli\jy} inside the primary beam \citep{2020:mertens}. To study the effect of DI-gain errors due to the absence of diffuse emission in the calibration sky model with simulations, however, such an elaborate model is not necessary, since the foreground power is dominated by the brightest point sources. Therefore, an NCP sky model 
with fewer components was created \citep{2024:Brackenhoff} in order reduce computational cost, which scales approximately linearly with the number of sources. The reduced sky model of the NCP 
consists of 684 unpolarized components with a flat spectrum and is shown in Figure \ref{fig:skymodel}. The model was created with \texttt{WSCLEAN} \citep{2014:offringa} on a small LOFAR dataset with a \qtyproduct{10 x 10}{\degree} field of view. Clean-components within the scale of the point spread function (PSF) were merged into single point sources. The flux densities in the sky model represent intrinsic flux densities and their location in the NCP field is shown in Figure \ref{fig:skymodel}. The cylindrically averaged power spectra of the full NCP model, consisting of 28778 components used for the real LOFAR data 
processing, and the 684-component model used in this paper agree within 10\% \citep{2024:Brackenhoff}. The sky model is stored as a text file in the "local sky model" (LSM) 
format, which is then input to the prediction of visibilities with \texttt{SAGECAL-CO} with application of the beam, which is discussed in Subsection \ref{subsec:beam}.

\subsubsection{Diffuse Galactic emission model}
Models of diffuse Galactic emission at MHz frequencies remain poorly constrained on small spatial and spectral scales, leading to significant model uncertainties. The most comprehensive all-sky map is still the Haslam map at \SI{408}{\mega\hertz} \citep{1981:haslamI, 1982:HaslamII}, which has an angular resolution of $\sim$\SI{1}{\degree}. A commonly used model in the literature is the Global Sky Model (GSM) \citep{2008:oliveira-costa} and its updated version, the improved GSM \citep{2017:zheng}, both of which are based on a principal component analysis and heavily influenced by CMB maps in the GHz frequency range. However, the GSM is not reliable on finer angular scales. 
Some lower-frequency measurements of diffuse emission exist from MWA, LWA1, and OVRO-LWA data \citep{2022:bryne, 2017:dowell, 2018:eastwood} at frequencies below \SI{200}{\mega\hertz}. 
The \SI{182}{\mega\hertz} diffuse emission map by \citet{2022:bryne} covers angular scales from 1 to 9 degrees but is limited to a single frequency without spectral information. 
The LWA1 Low Frequency Sky Survey \citep{2017:dowell} spans \SIrange{35}{80}{\mega\hertz} with angular scales of 2 to 4.7 degrees. \citet{2022:gehlot} modeled the diffuse Galactic emission at \SI{122}{\mega\hertz} 
around the NCP with the LOFAR-AARTFAAC system using multiscale CLEAN and shapelet decomposition. While the multiscale CLEAN method is able to model extended emission at intermediate angular scales of $\lesssim \SI{2.3}{\degree}$, 
it is sub-optimal for modeling larger scales of the order of a few degrees. The shapelet decomposition on the other hand is best suited to model larger angular scales.

We found that none of these models were suitable as a model of the diffuse Galactic foregrounds for the desired observational field, angular resolution and spectral information. 
We therefore describe the foregrounds as a Gaussian random field generated from an angular covariance function informed by the above-mentioned NCP data obtained with LOFAR-AARTFAAC. Spatially, the Gaussian random field does not resemble our Galaxy, but it provides a good approximation, as its underlying statistic, the angular correlation function, is similar. Nevertheless, for the purpose of studying the effects of unmodeled diffuse Galactic emission on calibration, this approach is sufficient

\begin{figure}
    \includegraphics[trim={0cm 0cm 1cm 6cm},clip, scale=0.26]{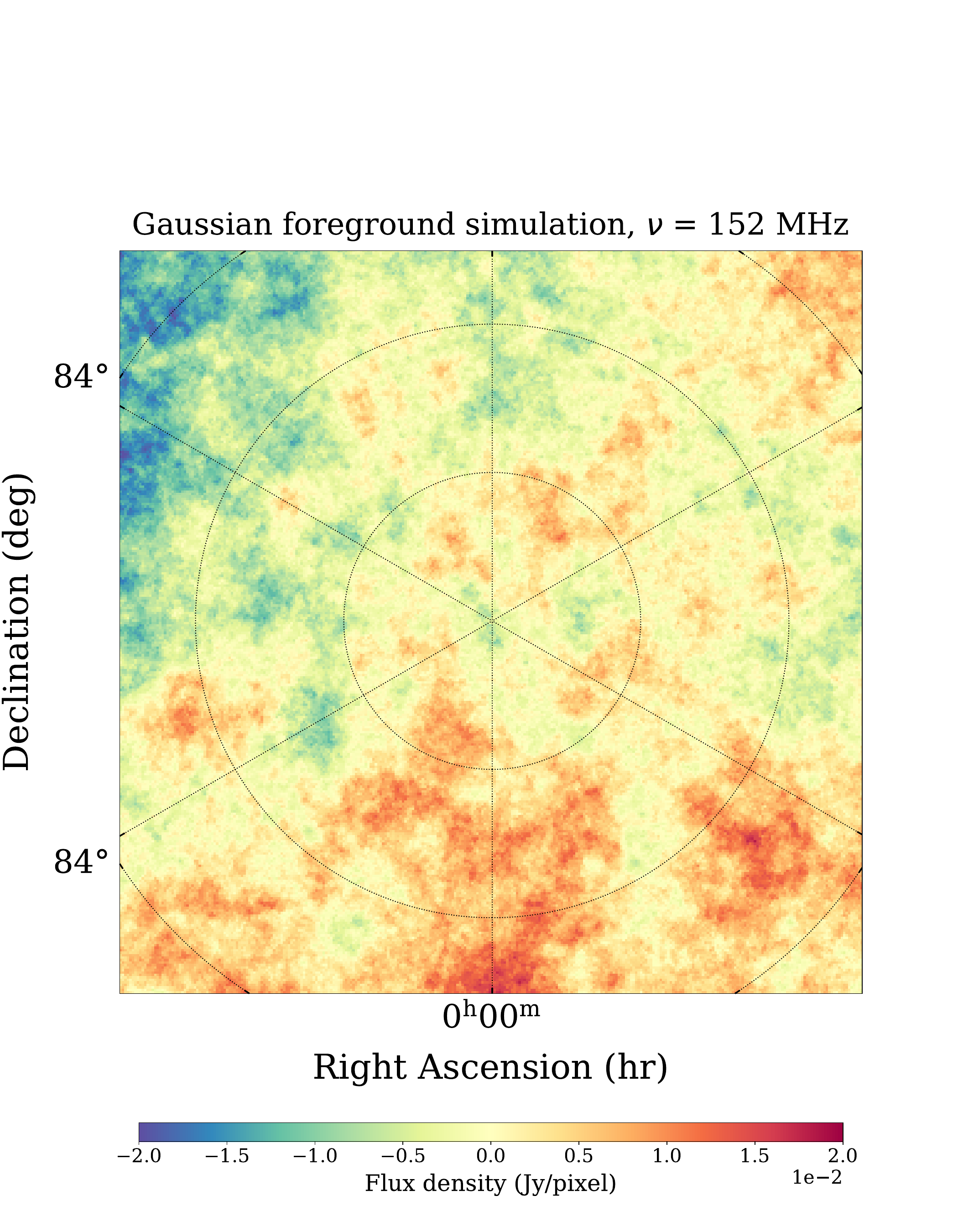}
    \caption{The diffuse Galactic emission is modeled as a Gaussian random field with an angular covariance informed by measurements of the NCP at \SI{122}{\mega\hertz} with LOFAR-AARTFAAC data and the Fan Region at \SI{150}{\mega\hertz} with the Westerbork telescope.}
    \label{fig:diffuse_emission}
\end{figure}

Our model of the diffuse Galactic synchrotron emission is based on \citet{2005:santos} and is described by an angular power spectrum as a function of angular scale $l$ and frequency $\nu$ of the form:
\begin{equation}
\label{eq:diffuse_emission}
C_l(\nu, \nu') = A \left( \frac{l}{l_0}\right)^{-\alpha} \left( \frac{\nu\nu'}{\nu_{0}^2}\right)^{-\beta} e^{-\frac{1}{2 \zeta_l^2} ln^2 (\nu / \nu')},
\end{equation}
where $\alpha$ is the power law index for the spatial spectrum, $\beta$ is the power law index for the frequency spectrum and $\zeta$ describes frequency correlation. For this work, we only consider unpolarized Galactic synchrotron emission. 
The parameters for this model are informed by \citet{2022:gehlot} and \citet{2009:bernadi}. \citet{2022:gehlot} measured the Galactic radio emission at \SI{122}{\mega\hertz} around the NCP with LOFAR-AARTFAAC at angular scales $20 \lesssim l \lesssim 200$ and found a brightness temperature variance of $\Delta^2_{l=180} = (145.64 \pm 13.61)K^2$ on angular scales of 1 degree. Using the relation $\Delta_l^2 \equiv l(l+1)C_l / 2 \pi$, we define the amplitude $A$ of the angular power spectrum model (Equation \ref{eq:diffuse_emission}) as the brightness temperature variance at $l_0 = 180$ and $\nu_0 = $ \SI{122}{\mega\hertz}. We find $A =$ \SI{0.0028}{\K^2}. 
The power-law index parameter for the spatial spectrum $\alpha$ is guided by measurements of the NCP with LOFAR-AARTFAAC by \citet{2022:gehlot} who 
found $\alpha = 2.0$ for $l \lesssim 200$ and measurements of the low Galactic latitude Fan region with the Westerbork synthesis radio telescope \citep{2009:bernadi2} who found $\alpha = 2.2$ for $l \lesssim 900$. We adopt $\alpha = 2.2$, for our model because the spatial spectrum in \citet{2009:bernadi2} was fitted over a broader multipole range. Detection of diffuse structure beyond $l \sim 900$ is limited by residual point sources, but we assume the power-law scaling holds to higher multipoles for modeling purposes. For the spectral index, we adopt $\beta = 2.55$ for frequencies between \SIrange{100}{200}{\mega\hertz} and high Galactic latitudes \citep{2008:rogers}. 
The spectral index of the angular power spectrum of the Galactic radio diffuse synchrotron emission typically steepens with increasing frequency corresponding to a softer electron spectrum \citep{2008:oliveira-costa}. For Galactic synchrotron emission, we adopt a frequency correlation parameter $\zeta = 4$, following \citet{2005:santos}.
We summarize the model parameters of the angular power spectrum in Table \ref{tab:galactic_model}.
\begin{table}
\begin{center}
\caption{Parameters for our diffuse Galactic emission angular power spectrum model}.
\label{tab:galactic_model}
\setlength\extrarowheight{1mm}
\begin{tabular}{c c c c c c}
 \hline \hline
 Component & Polarisation & A ($K^2$) & $\alpha$ & $\beta$ & $\zeta$ \\
\hline
 Galaxy & TT & 0.0028 & 2.2 & 2.55 & 4.0 \\ 
 \hline \hline
\end{tabular}
\tablefoot{The angular power spectrum model is given in Equation \ref{eq:diffuse_emission}. The parameters are based on the models of \citet{2005:santos} and informed by LOFAR-AARTFAAC data of the NCP \citep{2022:gehlot}. TT refers to the temperature-temperature (T-T) power spectrum.}
\end{center}
\end{table}
We simulate a full-sky Gaussian foreground \texttt{healpix} map modeled by the angular covariance defined in Equation \ref{eq:diffuse_emission}, using the model parameters from Table \ref{tab:galactic_model} and the Python package \texttt{cora}\footnote{\url{https://github.com/radiocosmology/cora}}. This simulation spans a frequency range of \SIrange{146.9}{159.6}{\mega\hertz}, corresponding to a redshift range of $z \sim 8.0 - 8.7$, one of the redshift bins analyzed in the LOFAR-EoR NCP data processing. We choose \texttt{healpix} \texttt{NSIDE} parameter of 2048, corresponding to an angular resolution of \SI{1.7}{\rm arcmin}. A \qtyproduct{10 x 10}{\degree} field of view is projected onto a 2D Cartesian grid with \SI{2}{\rm arcmin} resolution to eliminate nearest-neighbor interpolation effects. To avoid Gibbs ringing, the mean of each 2D image is subtracted. Finally, we convert brightness temperature $T$ to flux density in units of Jy/pixel, as required for predicting visibilities with \texttt{WSCLEAN}, using the Rayleigh-Jeans law
\begin{equation}
S = \frac{2 k_{\rm B} \Omega}{\lambda^2} T, 
\end{equation}
where $k_{\rm B}$ is the Boltzmann constant, $\Omega$ is the image pixel area, and the flux density $S$ is expressed in units of $10^{-26} \frac{\mathrm{W}}{\mathrm{m}^2 \mathrm{Hz}}$. The 2D images are then placed into \texttt{fits} files using the header information from images created from predicted visibilities of the NCP model, which includes information such as the pointing of the telescopes, the observing frequency, the point spread function. 

\subsubsection{Prediction of the beam}
\label{subsec:beam}
The prediction of the LOFAR-HBA beam was enabled using \texttt{SAGECAL-CO} during visibility simulation. To achieve a realistic representation of LOFAR-HBA observations of the sky, both the element beam and array factor were included. The element beam is the response of a single dipole antenna within a tile, with an instantaneous field of view of half the sky. The array factor models phasing up the 16 dual-polarization antennae to a tile beam, and further phasing of tiles into a station beam. The simulation also incorporates the effects of defunct and switched-off tiles, as recorded in the measurement set, to accurately represent the distorted station beam. The same beam model is also applied during data analysis, ensuring consistency between the simulated visibilities and the calibration/imaging steps, and therefore avoiding any discrepancies due to mismatched beam assumptions.

\subsubsection{Prediction of visibilities}
\label{sec:predict}
The prediction of visibilities for the point-source model is implemented with \texttt{SAGECAL-CO} which calculates the radio interferometer measurement equation (RIME) \citep{1996:hamaker, 2011:smirnov} for every single point source $k$ provided in the sky model file:
\begin{equation}
    \mathbf{V}_{ij}(t, \nu) = \sum_k \mathbf{J}_{i;k}(t, \nu) \mathbf{C}_{ij;k} (t, \nu)\mathbf{J}_{j;k}^{\dagger}(t, \nu).
\end{equation}
The visibility matrix $\mathbf{V}_{ij}(t, \nu)$ is a $2 \times 2$ complex matrix that represents the full-polarization cross-correlations between the signals from two beam-formed stations $i$ and $j$, as a function of time $t$ and frequency $\nu$. $\mathbf{J_i}$ is a $2 \times 2$ complex matrix called the Jones matrix describing the instrumental response of a radio interferometer to the incoming electric field. $\mathbf{C}_{ij;k}$ is the coherency matrix for each sky model component $k$ measured by baseline $ij$. It essentially describes the flux density of a compact source component in terms of the Stokes parameters $I, Q, U$ and $V$.

Visibilities were predicted using the $uvw$-coordinates and local sidereal time (LST) ranges from the LOFAR-HBA EoR observation targeting the NCP with a phase center at DEC = \SI{90.0}{\degree}. We simulated a total bandwidth of \SI{12.7}{\mega\hertz} between observing frequencies \SIrange{146.9}{159.6}{\mega\hertz} corresponding to the lowest redshift bin using all core and remote stations. The measurement set of the real LOFAR observation, used as a template in this study, provides all essential information about the interferometric array, including details of inactive tiles per station, allowing a realistic simulation of the time-dependent station beam. To reduce computation time, we limited the dataset size by selecting a 4-hour observation window while maintaining good $uv$-coverage. For the same reason, we adopted a time resolution of \SI{10}{\s} and a frequency resolution of \SI{195}{\kilo\hertz}. A summary of properties of the simulated observational dataset is presented in Table \ref{tab:simulated_dataset}. 

\begin{table}
\begin{center}
\caption{Properties of the simulated dataset based on real LOFAR-HBA observations of the NCP.}
\setlength\extrarowheight{1mm}
\begin{tabular}{l c}
 \hline \hline
 Parameter & Value \\
 \hline
 System & LOFAR-HBA \\
 Number of stations & 60 \\
 Pointing & DEC = \SI{90.0}{\degree}\\
 Duration & \SI{4}{\hour} \\
 Time resolution & \SI{10}{\s} \\
 Spectral bandwidth & \SIrange{146.9}{159.6}{\mega\hertz} \\
 Redshift range & $z \sim 8.0. - 8.7$ \\
 Spectral resolution & \SI{195}{\kilo\hertz} \\
 \hline \hline
\end{tabular}
\label{tab:simulated_dataset}
\end{center}
\end{table}

The simulation of the diffuse emission was performed with the image-domain gridder \citep{2018:vandertol, 2021:veenboer} of \texttt{WSCLEAN}, which enables the application of both the array factor and the primary beam. A $300 \times 300$ pixel image corresponds to simulating $\sim 10^{5}$ components with the standard RIME equation, which is computationally infeasible with our current computational resources. Gridding involves applying a fast Fourier transform to the image, where instead of calculating the transform at every possible point in the $uv$-plane it is computed only at regularly spaced grid points. This approach significantly speeds up the prediction of visibilities for diffuse emission, compared to using a direct Fourier transform.

\subsubsection{Simulation of thermal noise}
\label{subsec:noise}

Next, we simulated thermal noise according to the radiometer equation:
\begin{equation}
\sigma = \frac{\textrm{SEFD}}{\sqrt{2 \Delta t \Delta \nu}},
\end{equation}
The system equivalent flux density (SEFD) is a measure of the sensitivity of a radio telescope system, $\Delta t$ is the integration time and $\Delta \nu$ is the spectral resolution. The SEFD was measured for each station as a function of frequency and tabulated \citep{2013:van_Haarlem}. Thermal noise was generated by drawing from a Gaussian distribution given by $\sigma$ and a mean of zero and added to the visibility matrix using the LOFAR Simulation Tool \texttt{losito}.

\subsection{LOFAR-EoR data processing pipeline}
In this section, we outline the key steps of the LOFAR-EoR data processing pipeline for the NCP deep field \citep{2020:mertens} relevant to this study. These include DI-gain calibration of visibilities using \texttt{SAGECAL-CO}, imaging the gain-corrected gridded visibilities with \texttt{WSCLEAN}, and power spectrum estimation using \texttt{pspipe}.

\subsubsection{Direction-independent gain calibration}
\label{sec:DI_gain_cal}
DI-gain calibration refers to the correction of complex gain variations that are common across the entire field of view of the radio interferometer and therefore do not vary with direction on the sky. Key sources of direction-independent errors include: (1) complex gain variations in the analog signal chain due to temperature fluctuations, cable reflections, etc.; (2) clock and time synchronization errors; (3) ionospheric effects on short time scales and (4) differences between the sky model and the true sky, along with beam errors, which can lead to inaccuracies in calibration, contributing to an overall mismatch between the expected and the observed sky . DI-gain calibration aims to correct these types of errors by using a catalog of bright sources, the sky model. The resulting gain solutions are then applied uniformly across the entire field of view, correcting for the average flux density of the field.

Since the brightest source in the NCP field, 3C 61.1, is close to the first null of the primary beam of the station, DI-gain calibration for observational data is actually performed in two directions to separate the strong direction-dependent effects for this source. For our forward simulations, we decided to stick to calibrating to a single direction, since our simulated visibilities are not corrupted by any gain errors other than the ones introduced by calibrating with a sky model omitting diffuse Galactic emission. We also found no significant difference in the gain solutions when calibrating in one direction compared to two directions in our simulations.

DI-gain calibration is performed using \texttt{SAGECAL-CO}, which employs consensus optimization \citep{2011:boyd} to iteratively adjust station gain solutions, driving them towards spectrally smooth functions that are also optimized. The key idea behind consensus optimization is to have different agents work on separate parts of the problem while ensuring that their solutions converge to a common consensus. At each iteration, solutions that deviate from a frequency-smooth prior are penalized by a regularization term, the strength of which is determined by a parameter $\rho$, which is then updated. Eventually this process should converge to the prior. For a more detailed and mathematical description of the \texttt{SAGECAL-CO} algorithm, we refer the readers to \citet{2015:yatawatta, 2016:yatawatta}.

Improving DI-gain calibration has been a major focus of reprocessing the LOFAR NCP data over the past year. This was accomplished by employing a two-step approach to minimize the number of free parameters in the DI-gain calibration, leading to a reduction in gain errors on small spectral scales that could otherwise affect or mimic the \SI{21}{\centi\metre} signal. Delay spectra of the normalized gain solutions showed a decrease in power by two to three orders of magnitude when using the two-step approach with regularized gains, compared to the one-step approach with unregularized gains \citep{2025:mertens}. For our simulations, we followed the same two-step DI-gain calibration scheme:
\begin{itemize}
    \item Step 1: Correction of smooth spectral and fast temporal gain variations
    \item Step 2: Correction of fast varying bandpass and slow temporal gain variations
\end{itemize}

For the first DI-step, we used a third-order Bernstein polynomial over the \SI{12.7}{\mega\hertz} bandwidth, consistent with the analysis of LOFAR-EoR NCP data \citep{2020:mertens}. The optimal regularization parameter was found by running DI-gain calibration on forward simulated noisy visibility data as a function of regularization parameter with a fixed number of iterations and quantifying the variance of the sub-band differenced gain solutions. A clear minimum was found at $\rho$ = 200, which verifies the results by \citet{2016:yatawatta} and \citet{2020:mertens}. Low or no regularization will over-fit the data, resulting in signal suppression on the smallest baselines where we are most sensitive to the \SI{21}{\centi\metre} signal \citep{2017:patil, 2022:mevius}. Calibration was performed on a spectral and time solution interval of \SI{195.3}{\kilo\hertz} (one sub-band) and \SI{30}{\second}, respectively. We used a total of 40 ADMM (Alternating Direction Method of Multipliers \citep{2011:boyd}) iterations during the optimization, which was found to be sufficient to achieve the required convergence \citep{2022:mevius}. Baselines between $50 - 5000~\lambda$ were used.

In the second DI-gain calibration step, we solved for the fast frequency-varying bandpass response of the stations, which in real data, is caused by filter amplifiers in the signal chain and coaxial cable reflections between receivers and tiles, but does not vary with time. These effects, however, are not present in the simulated data. Nonetheless, for consistency with real data processing, we apply the same DI-gain calibration procedure to the simulations. Unregularized DI-gain solutions were found on a per-frequency-subband basis. We set the time solution interval to \SI{4}{\hour}, which corresponds to the duration of the simulated observation and is also the maximum amount of data that can be processed by a single NVIDEA K40 GPU on the supercluster \texttt{DAWN}, the high-performance computing (HPC) system currently used for analysis of real LOFAR-EoR observations and simulations. Again, baselines between $50 - 5000~\lambda$ were used. The gain solutions from both DI steps were applied sequentially to the simulated visibilities for the entire field of view.

\subsubsection{Imaging}
The gain-corrected visibilities were gridded and each sub-band was imaged independently using \texttt{WSCLEAN} \citep{2014:offringa}, creating an ($l$, $m$, $\nu$) image cube. We followed the same gridding procedure as in the LOFAR NCP data processing \citep{2020:mertens} with unit weighting during gridding and a Kaiser-Bessel anti-aliasing filter with a kernel size of 15 pixels, an oversampling factor of 4096 and 32 w-layers. Additionally, we selected baselines between $30 - 250~\lambda$, \SI{0.5}{\rm arcmin} resolution and $1024 \times 1024$ pixels. These settings ensured that any systematic effects due to gridding are subdominant to the expected \SI{21}{\centi\metre} signal and thermal noise (see Figure 8 in \citet{2019a:offringa} and Figure 5 in \citet{2019b:offringa}). 
 
Stokes $I$ and $V$ images, in units of Jy/PSF, as well as the point-spread function (PSF) maps were created with natural weighting. To estimate the thermal noise variance in the data, images were produced from gridded time-differenced visibilities between even and odd (i.e., alternating) time samples. Different sub-bands were then combined to form image cubes with a field of view of \qtyproduct{8.5 x 8.5}{\degree} and \SI{0.5}{\rm arcmin} resolution. To ensure that the analysis focused on the most sensitive part of the primary beam (with a full width at half maximum (FWHM) of \SI{4.1}{^\circ} at \SI{150}{\mega\hertz}), image cubes were trimmed using a Tukey spatial filter with a diameter of \SI{4}{^\circ}. This aligns with our approach to real data.

\subsubsection{Power spectrum estimation}
\label{subsubsec:ps_estimation}
The power spectrum estimation was performed using the Python package \texttt{pspipe}\footnote{Power Spectra Generation pipeline for CD/EoR experiments \url{https://gitlab.com/flomertens/pspipe/}} and described in detail by \citet{2020:mertens}. The cylindrically averaged power spectrum $P(\mathbf{k})$ as a function of wavenumber $\mathbf{k}$ is defined as: 

\begin{equation}
    P(k_{\perp}, k_{\parallel}) = \frac{X^2 Y}{\Omega_{PB} B} \left< | \hat{V}(\mathbf{u}, \tau) |^2 \right>,
\end{equation}
where $\hat{V}(\mathbf{u}, \tau)$ is the Fourier transform in the frequency directions, $\Omega_{\rm PB}$ is the integral of the primary beam gain squared over solid angle, $B$ is the frequency bandwidth and X and Y are conversion factors from angle and frequency to comoving distance. The expectation values indicate averaging over baselines within a ($k_{\perp}$, $k_{\parallel}$) power spectrum bin.

The Fourier modes $k_{\perp}$ and $k_{\parallel}$ are wave numbers in units of inverse comoving distance and given by \citet{2004:morales}:
\begin{equation}
    k_{\perp} = \frac{2\pi |\mathbf{u} |}{D_M(z)}, \quad k_{\parallel} = \frac{2\pi \nu_{21} H(z)}{c (1+z)^2} \tau, \quad k = \sqrt{k_{\perp}^2 + k_{\parallel}^2},
\end{equation}
where $D_M(z)$ is the transverse co-moving distance, $\nu_{21}$ is the frequency of the hyperfine transition of neutral hydrogen, and $H(z)$ is the Hubble parameter.

\section{Direction-independent gain calibration errors due to unmodeled diffuse Galactic emission}
\label{sec:calibration_errors}
As described in Section \ref{sec:DI_gain_cal}, gain calibration is done using a non-linear optimization method named \texttt{SAGECAL-CO}. Gain solutions are obtained through an iterative process, as the minimization of the Lagrangian lacks a closed-form solution. Consequently, deriving an analytic formalism for gain errors is challenging. While some approaches to linearizing the problem exist in the literature \citep{2014:Grobler, 2016:Wijnholds}, our focus is on providing a qualitative description of the gain calibration process in the presence of unmodeled Galactic diffuse emission, aiming to build intuition for the reader.

For the purposes of this paper we only consider unpolarized foreground emission, since this is the most dominant component, and set $Q, U, V = 0$. Furthermore, we assume that unpolarized foreground emission consists of a point source and a Galactic diffuse component, denoted by the coherency matrix $\mathbf{C}^{\rm P}$ and $\mathbf{C}^{\rm D}$, respectively. We define the true visibilities as the total foreground emission measured by the interferometer. They represent the "true" data:
\begin{equation}
\label{eq:V_true}
    \mathbf{V}_{ij}^{\rm true} (t, \nu) = \sum_k \mathbf{J}_{i;k} (t, \nu) \Bigl( \mathbf{C}_{ij;k}^{\rm P}(t, \nu) + \mathbf{C}_{ij;k}^{\rm D}(t, \nu) \Bigr) \mathbf{J}_{j;k}^{\dagger}(t, \nu) + \mathbf{N}_{ij},
\end{equation}
where $\mathbf{J}_{i;k}$ and $\mathbf{J}_{j;k}$ are the true Jones matrices (gains) of stations $i$ and $j$ in direction $k$, and $\mathbf{N}_{ij}$ represents the additive noise. During gain calibration, \texttt{SAGECAL-CO} uses a sky model to compute model visibilities, which are updated at each iteration based on the current gain estimates \citep{2015:yatawatta}. In this study, we assess calibration errors introduced by unmodeled diffuse Galactic emission. To quantify their effect on the gain solutions, we simulate visibilities that include both point sources and diffuse Galactic emission (PSDE), but calibrate them using a point-source-only (PSO) sky model. We refer to this setup as mismatched sky model calibration, since the sky models used in the forward simulation and calibration differ.

Since the goal is to minimize the difference between data and model, some of the power from the diffuse emission will be absorbed into the final gain estimates $\mathbf{\tilde{J}}_i$. A previous study \citep{2016:patil} demonstrated that when using all baselines, polarized diffuse emission can be absorbed into the direction-dependent gain solutions. This effect occurs because the point spread function (PSF) adapts to compensate for the missing diffuse sky model components, while largely preserving the positions and flux densities of discrete point sources.
DI-gain calibration means correcting for $\mathbf{\tilde{J}}_i$ by applying the inverse of the Jones matrix to retrieve the intrinsic flux density of the sky. Due to the omission of diffuse Galactic emission in the calibration sky model, the estimated gains deviate from the true gains, such that $\mathbf{\tilde{J}}_i^{-1} \mathbf{J}_i \neq \mathbb{1}$. DI-gains are applied directly to the simulated data, hence the DI-gain calibrated visibilities for every time step $t$ and frequency $\nu$ are:
\begin{equation}
\label{eq:V_cal}
    \mathbf{V}_{ij}^{\rm DI-cal} = \mathbf{\tilde{J}}_i^{-1} \left[ \mathbf{J}_i \left( \mathbf{C}_{ij}^{\rm P} + \mathbf{C}_{ij}^{\rm D} \right) \mathbf{J}_j^{\dagger} + \mathbf{N}_{ij} \right] \mathbf{\tilde{J}}_j^{\dagger -1}.
\end{equation}
We have dropped the direction index $k$, since we consider calibration only in one direction. As mentioned in Section \ref{sec:DI_gain_cal}, DI-gain calibration is performed in two steps: (1) correcting smooth spectral variations and fast time-varying fluctuations, and (2) correcting fast-varying band-pass effects and slow time-varying variations (bandpass calibration). Hence, our gain estimates are a multiplication of the Jones matrices of the two calibration steps: $\mathbf{\tilde{J}}_i = \mathbf{\tilde{J}}^{\rm DI1}_i \mathbf{\tilde{J}}^{\rm DI2}_i$. All results presented in this paper refer to the combined Jones matrix from DI-gain calibration steps 1 and 2.

In order to compare the estimated gains $\mathbf{\tilde{J}}_i$ affected by unmodeled diffuse emission to unaffected gains, we performed a control simulation where visibilities were forward-simulated using only the point-source catalog, hence:

\begin{equation}
    \mathbf{V}_{ij}^{\rm true}(t, \nu) = \sum_k\mathbf{J}_{i;k}(t, \nu)\mathbf{C}_{ij}^{\rm P}(t, \nu) \mathbf{J}_j^{\dagger}(t, \nu) + \mathbf{N}_{ij}.
\end{equation}

We then performed DI-gain calibration using a PSO sky model. As this scenario assumes perfect knowledge of the sky model and the instrument, except for thermal noise, we refer to it as matched point-source only (PSO) calibration or the ground truth throughout the paper. Under these idealized conditions, the true gains are expected to be close to unity.

\section{Results}
\label{sec:results}
This section analyzes the effects of DI-gain calibration errors due to the missing diffuse Galactic emission model. We first examine the gain solutions themselves, then assess the impact of these calibration errors in image space, and finally evaluate their effect on the 2D cylindrically averaged power spectra in comparison to thermal noise. Ultimately, we assess the extent to which these errors impact the detectability of the cosmological \SI{21}{\centi\meter} signal in the power spectrum.

\subsection{Gain solution analysis}
There are a few questions we would like to
address:

\begin{enumerate}
\item Is there a bias in the gain solutions when calibrating with
a sky model that is omitting diffuse Galactic emission, and by what percentage do they
deviate from unity compared to the ground truth?
\item Does the variance in the gain solutions increase when diffuse
Galactic emission is unmodeled in the calibration process?
\item Is the fractional change in power in the gain solutions dependent on baseline length, given that diffuse Galactic emission
is mainly detected with shorter baselines?
\end{enumerate}

As described in Section \ref{sec:DI_gain_cal}, for DI-gain calibration we used baselines in the range of $50 - 5000~\lambda$, a frequency regularization parameter of $\rho = 200$ and a time solution interval of \SI{30}{\s}. Since the baseline cut is in units of $\lambda$, some baselines including LOFAR remote stations have frequency channels for which the corresponding ($u,v$)-positions fall inside and outside the $5000~\lambda$ limit. The gain solutions for these stations were obtained from a subset of the correlation products and are therefore less constrained and noisier. In the future, LOFAR data processing will move to a baseline cut in units of meters. 

\begin{figure*}
    \centering
    \includegraphics[trim={2cm 1cm 1cm 2cm},clip, scale=0.4]{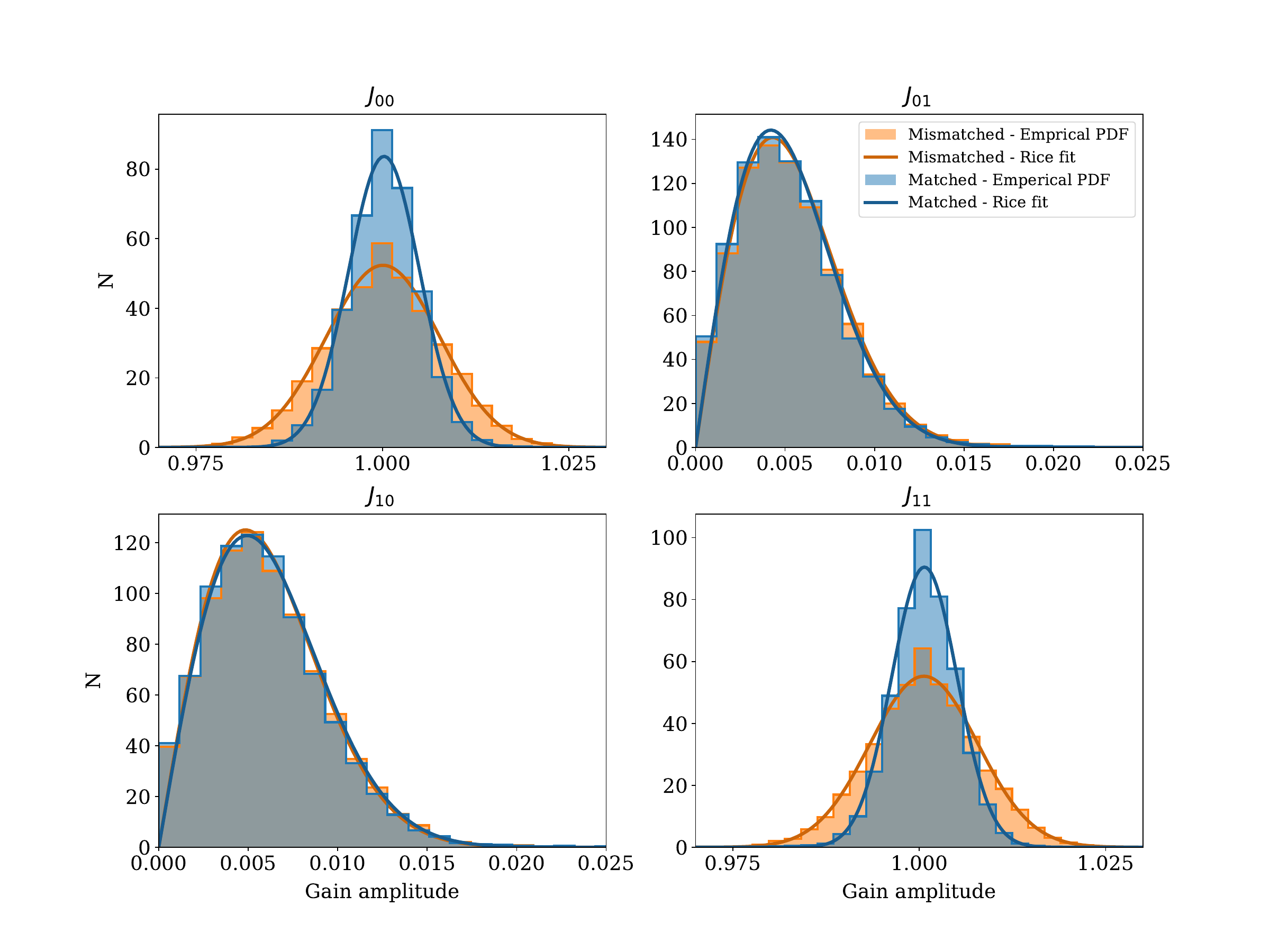}
    \caption{Normalized histograms of the gain amplitudes at \SI{152}{\mega\hertz} for all stations and time solutions. Solid lines show Rice distribution fits for each calibration scenario. Each panel represents a matrix element of the Jones matrix. The gains are referenced to the core station CS001HBA1 by multiplying them with the inverse gain matrix of CS001HBA1. The blue histogram shows the empirical PDF of gain amplitude solutions for the ground-truth case, obtained by forward-simulating and calibrating with a PSO sky model. The orange histogram shows the empirical PDF for gain solutions from simulations using a PSDE sky model but calibrated with a PSO model, thus omitting diffuse Galactic emission during DI-gain calibration, as described in the text.}
    \label{fig:gain_hist}
\end{figure*}

In Figure \ref{fig:gain_hist}, we show the normalized histograms of gain amplitude solutions at \SI{152}{\mega\hertz} for all stations and times, representing empirical probability density functions (PDFs). This frequency is representative for mid-band frequencies, whereas frequencies at the edge of the bandwidth generally have wider distributions due to the functional shape of the third-order Bernstein polynomial. The gains were referenced to the Jones matrix of the core station CS001HBA1 by multiplying the gain matrices with the inverse gain matrix of CS001HBA1. Each subplot represents the amplitude of the complex gain solutions for an element of the Jones matrix. $\mathbf{J}_{00}$ represents the response of the x-polarization feed to the x-component of the electric field $E_x$, $\mathbf{J}_{01}$ represents the response of the x-polarization feed to the y-component of the electric field $E_y$ and so forth. The sky model was forward simulated as unpolarized, so the off-diagonal elements are expected to be zero, except for instrumental polarization leakage caused by non-orthogonal feeds. Blue shows the gain amplitude solutions for the ground truth case, obtained by forward-simulating and calibrating with a PSO sky model. Orange indicates the gain amplitude solutions obtained from a mismatched sky model calibration, where visibilities are forward-predicted using both point sources and diffuse emission (see Equation \ref{eq:V_true}), but calibration is performed using only the PSO catalog. Overlaid are the Rice distribution fits for both calibration scenarios, obtained via MCMC sampling of the posterior distribution for the noncentrality parameter, as well as the location and scale parameters. The Rice distribution fit for $J_{00}$ reveals that the noncentrality parameter decreases from 9.0 in the matched PSO sky model calibration to about 8.1 in the mismatched case, indicating reduced signal coherence when diffuse Galactic emission is omitted. Similarly, the scale parameter $\sigma$ increases from roughly 0.0048 to 0.0076, reflecting greater scatter in the gain amplitude solutions under the mismatched sky model calibration. The location parameter $\mu$ shows a slight decrease from about 0.96 to 0.94, suggesting a minor systematic shift in the gain amplitude distributions. We find similar results for $J_{11}$.
 
To summarize, the standard deviation on the diagonal of the Jones matrix increases by approximately a factor of two when using a mismatched sky model calibration. This means that, on average, the power of the diffuse emission is absorbed or redistributed into the gain solutions, making them noisier when calibrating with a PSO sky model and introducing a systematic error. Note that this normalized histogram is over all stations and time samples and does not highlight any baseline dependent effects.

In Figure \ref{fig:gain_mat1}, we show the effect of neglecting the Galactic diffuse emission in the DI-gain calibration process on the gain visibility matrix as a function of baseline length. The top panel shows the baseline lengths corresponding to the upper triangle of the $N_{\rm station}^2$ visibility matrix. The first 12 core stations constitute the heart of the core, named the "Superterp" (ST), and reside on a \SI{320}{\m} diameter island. They create the shortest baselines of the LOFAR interferometer most sensitive to diffuse emission, apart from the intra-core station baselines (first off-diagonal matrix elements). The remaining 36 core stations (CS) form baselines extending up to approximately $2000~\lambda$, followed by 14 remote stations (RS). While CS and RS are often distinguished by baseline length, a more critical distinction for calibration is that core stations share a centralized clock enabling coherent beamforming, whereas remote stations use independent, GPS-disciplined rubidium clocks and therefore require explicit correction for clock offsets and drift, particularly important when working with real observational data. Furthermore, to approximate the beam characteristics of core stations, LOFAR remote stations deactivate 24 of their 48 tiles. The resulting beam pattern still differs slightly from that of a true core station due to residual electromagnetic coupling between the active inner tiles and the powered-off outer tiles \citep{2025:mertens}.

Baselines $< 50~\lambda$ and $> 5000~\lambda$ at frequency \SI{152}{\mega\hertz} are masked in grey since they are not considered during DI-gain calibration. In the middle panel we show the amplitude of the gain visibility matrix averaged over time solutions for the ground truth case, obtained by forward-simulating and calibrating with a PSO sky model for frequency \SI{152}{\mega\hertz}. We created an upper triangle gain visibility matrix by multiplying the station gain $\mathbf{J}_i$ with the corresponding Hermitian $\mathbf{J}_j^{\dagger}$ for baseline $ij$ and computing the absolute value of the first element of the Jones matrix: $|\mathbf{J}_i \mathbf{J}_j^{\dagger}|_{00}$. Variations in the gain products are of the order of $\sim 0.5\%$, likely due to thermal noise variations in the visibilities. RS with baselines near the $5000~\lambda$ cut have less constrained gain solutions due to missing data, resulting in high noise levels. As a result, their gains deviate by more than 1\% from unity. The bottom panel indicates $|\mathbf{J}_i \mathbf{J}_j^{\dagger}|_{00}$ when forward-simulating a PSDE sky model and calibrating with a PSO sky model, which omits the diffuse Galactic emission in the calibration process. Variations increase from 0.5\% to 1\% in specific gain visibilities involving the Superterp and nearby core stations, indicating that diffuse Galactic emission is absorbed into the gain solutions as a function of baseline length, with shorter baselines showing a stronger effect.

\begin{figure}
    \centering
    \includegraphics[trim={10cm 6.5cm 3cm 7cm}, clip, scale=0.36]{202507_mean_gain_amp_f152_J00_DI_step1_step2_all_v2_incl_pb.png}
    \caption{Top: Baseline lengths for the upper triangle of the visibility matrix. ST denotes Superterp core stations, CS core stations, and RS remote stations; these labels act as proxies for baseline lengths. Baselines $< 50~\lambda$ and $> 5000~\lambda$ at an observing frequency of \SI{152}{\mega\hertz} are masked in grey.  Middle: Absolute value of the time-averaged gain products for the upper triangle of the visibility matrix in the ground truth scenario. Bottom: Same as the middle panel, but forward-simulating with a PSDE sky model and calibrating with a PSO sky model, thereby neglecting diffuse Galactic emission during DI-gain calibration.}
    \label{fig:gain_mat1}
\end{figure}

\subsection{Analysis in image space}

\begin{figure*}[h]
    \includegraphics[trim={4cm 2cm 5cm 8cm},clip=true, scale=0.26]{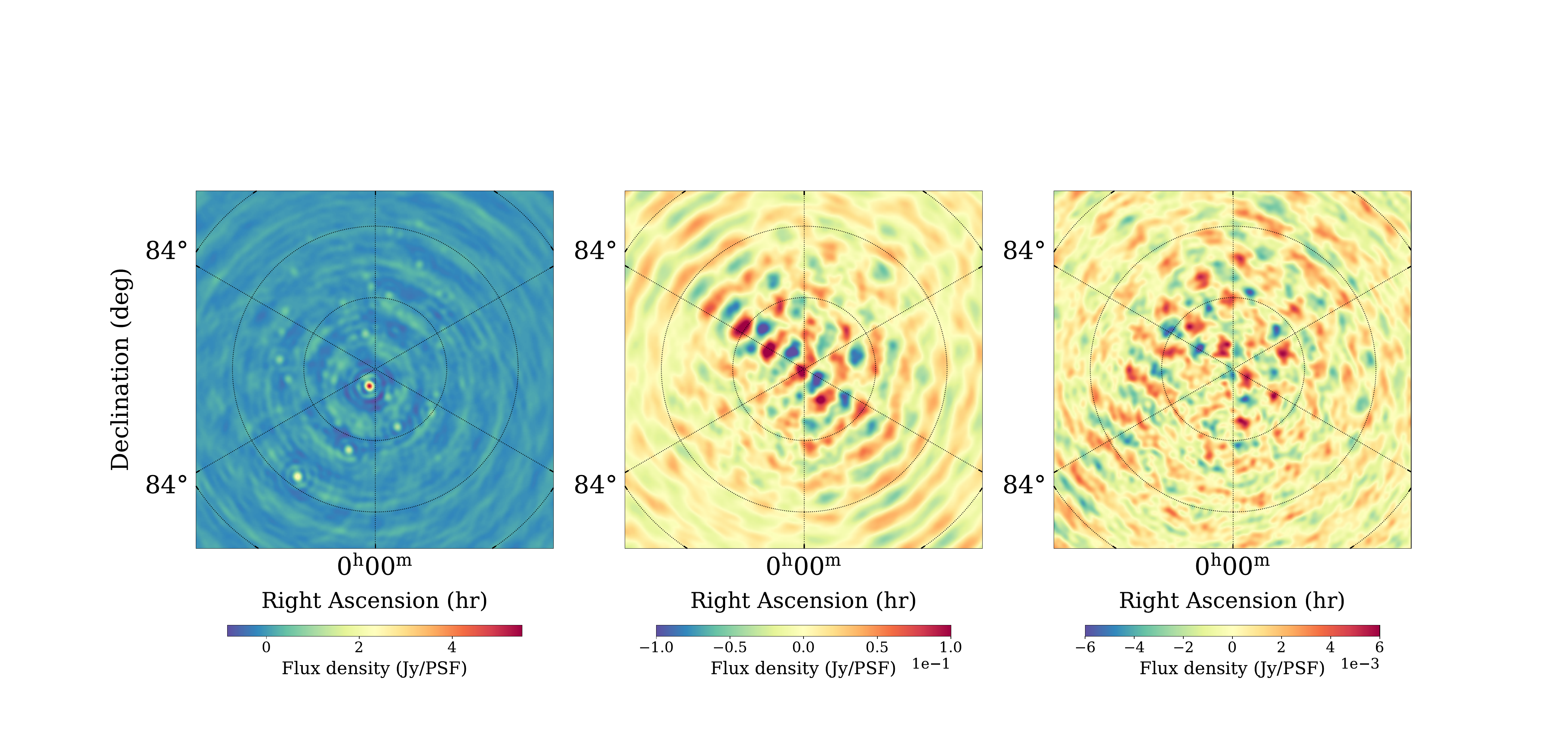}
    \caption{Left panel: Image of the true sky including point sources and diffuse emission. All images are created at frequency \SI{152}{\mega\hertz} with a minimum and maximum baseline distance of 50$\lambda$ - 250 $\lambda$. Middle panel: The diffuse emission with a \SI{4}{\hour} track point spread function of LOFAR-HBA applied. Right panel: The residual emission is obtained by subtracting the image created before DI-gain calibration from the one created afterward.}
    \label{fig:diff_img_50_250}
\end{figure*}

To show the effect of calibration errors due to the missing diffuse Galactic emission component in image space, we created images from the true (before DI-gain calibration) and DI-gain calibrated visibilities with a minimum and maximum baseline distance of $50 - 250~\lambda$. These are similar baselines that are used in LOFAR's EoR \SI{21}{\centi\m} power spectrum analysis. All images were created using \texttt{WSCLEAN} with an angular resolution of \SI{2}{\rm arcmin}, $10^\circ \times 10^\circ$ FoV and uniform weighting. The units are in Jy/PSF. Figure \ref{fig:diff_img_50_250} shows the full sky model (point sources and diffuse emission) on the left panel, and the diffuse emission as seen by the LOFAR-HBA instrument in the middle panel. The diffuse emission is at least an order of magnitude fainter than the point sources for baselines $50 - 250 \lambda$. The imprint of the primary beam is clearly visible, suppressing structures beyond the FWHM of $\sim$ \SI{4}{^\circ}. The right panel shows the residual emission by differencing the image of the DI-gain calibrated sky from the true sky. We see an RMS of $\sim$ \SI{6e-3}{\jy}/PSF which is an order of magnitude less than the RMS from the modeled diffuse emission. For shorter baselines the gain errors are between $0.5\% - 1\%$, which corresponds to flux density fluctuations of $\sim$ \SI{5e-3}{\jy} to \SI{1e-2}{\jy} for a $\SI{1}{\jy}$ source, which is similar to the level of residual emission we observe in the right panel of Figure \ref{fig:diff_img_50_250}. 

To further illustrate the impact of diffuse Galactic emission during DI-gain calibration, we compare residual emission in image space between two scenarios: (1) matched PSO calibration, where both the simulated sky and calibration model contain only point sources (left panel of Figure \ref{fig:diff_img_ps} in Appendix \ref{sec:appendix_c}); and (2) mismatched sky model calibration, where the simulated sky includes both point sources and diffuse emission (PSDE), but calibration is performed using only the PSO model (right panel of Figure \ref{fig:diff_img_ps} in Appendix \ref{sec:appendix_c}, identical to the right panel of Figure \ref{fig:diff_img_50_250}). In both cases, the residual emission is computed by subtracting the DI-gain-calibrated sky from the true sky. This comparison clearly shows that diffuse Galactic emission is absorbed into the DI-gain solutions for mismatched sky model calibration and appears as residual emission in image space. Since the inverse of the gains is applied directly to the visibilities this leads to a propagation of these errors to subsequent data analysis steps such as power spectrum estimation, which is discussed next.

\subsection{Fiducial power spectra}
\begin{figure*}
    \centering
    \includegraphics[trim={1cm 0 1cm 1cm},clip, scale=0.35]{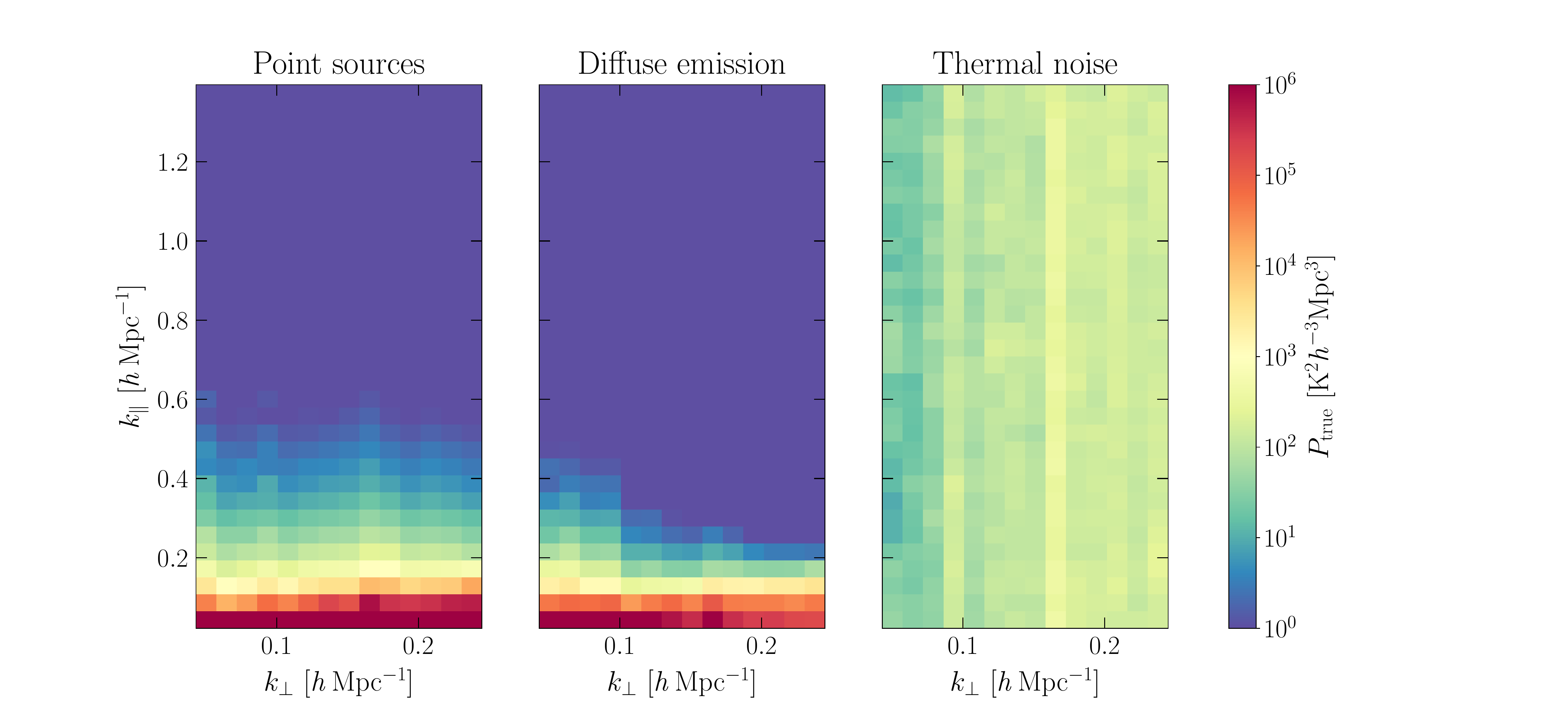}
    \caption{The fiducial 2D cylindrically averaged power spectra of simulated data from a point source model (left panel), from a diffuse Galactic emission model (middle panel) and from thermal noise (right panel).}
    \label{fig:ps_de_n}
\end{figure*}

Figure \ref{fig:ps_de_n} shows the fiducial power spectra of the different components of our simulated data, namely the point sources (left panel), the diffuse Galactic emission (middle panel) and the thermal noise (right panel). Spectrally smooth foregrounds dominate the lowest  $k_{\parallel}$-bins. The point source power spectrum shows higher power at larger $k_{\perp}$ because the power spectrum of point sources remains flat with respect to angular scale (which is inversely proportional to baseline length), whereas the power spectrum of diffuse emission decreases as the angular scale becomes smaller. The spread of foreground power into higher $k_{\parallel}$-bins at increasing $k_{\perp}$, known as the "foreground wedge", is caused by the instrument's chromaticity. The noise power spectrum is flat as a function of $k_{\parallel}$ for fixed $k_{\perp}$. Power spectrum bins associated with $uv$-bins that contain less visibility data exhibit higher noise power.

\subsection{Matched point-source-only calibration}
\label{subsec:complete_skymodel_cal}

In a control simulation, we generated simulated data that included only a point-source sky, without any diffuse Galactic emission. We added thermal noise to the simulation, creating ten different versions of this data, each with a unique realization of thermal noise drawn from the same distribution. Then, we calibrated these visibilities using a model that matched the point-source sky (i.e., the sky model used for calibration was complete and accurate). This process produced ten correctly calibrated power spectra, each representing the simulated data under a different thermal noise realization.

In this setup, any bias or variance in the power spectrum arises solely from thermal noise variations. To estimate the systematic bias, $b^{\rm sys}$, we calculated the mean residuals of the ten estimated power spectra relative to the fiducial true power spectrum using the power spectrum estimator discussed in Section \ref{subsubsec:ps_estimation}: 
\begin{equation}
\label{eq:bsys}
b^{\rm sys} = \left< \Delta P \right> = \left< P_i - \left< P_{\rm true} \right>\right>.
\end{equation}
Here, $P_i$ represents the calibrated power spectrum, while $\left< P_{\rm true} \right>$ represents the fiducial true power spectrum, derived by averaging ten power spectra generated from true visibilities $V_{ij}^{\rm true}$ under variation of thermal noise. The angle brackets $\langle ...\rangle$ denote ensemble averages over the ten realizations. We will define several configurations for our simulations in this study:

\begin{itemize}
    \item Setup 1 is the ground truth and control simulation. $\left< \Delta P_{N_i}^{\rm DI-cal} \right>$ is the estimated expectation value of the power spectrum residuals, when DI-gain calibration is performed with a matched sky model and the thermal noise realization is varied. The fiducial power spectrum $\left< P_{\rm true} \right> = \left<P_{\rm P} + P_{N_i} \right>$ is the ensemble average of power spectra generated from true visibilities with different thermal noise realizations.
\end{itemize}
Since thermal noise and solver noise\footnote{This term refers to the additional noise introduced by the random initialization of parameters in \texttt{SAGECAL-CO} for each calibration run, resulting in slight variations in gain solutions after a finite number of optimization iterations \citep{2022:mevius}.} is incoherent between realizations, we expect $\left< \Delta P_{N_i}^{\rm DI-cal} \right>$ to approach zero as the number of realizations increases, except for in the lowest $k_{\parallel}$-bins, where the foregrounds are located. We show the power spectrum residuals of Setup 1 as a fraction of the expected thermal noise in the left panel of Figure \ref{fig:delta_p}. It should be noted, that even when we calibrate with a matched sky model, there is a systematic bias in the lowest $k_{\parallel}$-bins simply due to thermal noise variations. Thermal noise introduces slight inaccuracies in gain calibration, subsequently modulating the foregrounds. A key question arises: if the impact on the gains is random, why does it still lead to a systematic offset that results in an overall increase in power in the lower $k_{\parallel}$-bins? When constructing the power spectrum, visibilities within a single $uv$-cell are averaged and then squared. DI-gain corrected visibilities are modified by two estimated Jones matrices, $\tilde{\mathbf{J}}_i$ and $\tilde{\textbf{J}}^{\dagger}_j$, following Equation \ref{eq:V_cal}. If both $\tilde{\mathbf{J}}_i$ and $\tilde{\textbf{J}}^{\dagger}_j$ contain Gaussian noise, the expectation value of Equation \ref{eq:V_cal} becomes biased, leading to an excess in power. The bias is confined to the lowest $k_{\parallel}$-bin, due to the frequency-smooth constraint during DI-gain calibration.

\begin{figure*}
    \centering
    \includegraphics[trim={1cm 0 3cm 1.5cm},clip=true, scale=0.35]{202507_delta_p_over_n_seismic.png}
    \caption{The systematic bias as a fraction of the thermal noise for a simulation with matched sky model DI-gain calibration (forward-simulation and calibration with a PSO sky model) in the left panel (Setup 1) and with a mismatched sky model DI-gain calibration (forward-simulation with a PSDE sky model and calibration with a PSO sky model, which is missing the diffuse Galactic foreground component), in the middle panel and right panel (Setup 2 and 3). The systematic bias is estimated from ten power spectra under variation of thermal noise (Setup 1), diffuse foregrounds (Setup 2), thermal noise and foregrounds (Setup 3), respectively, in the visibility simulation. Blue regions indicate that the systematic bias is below the estimated thermal noise and red regions above.}
    \label{fig:delta_p}
\end{figure*}

\subsection{Mismatched sky model calibration}
\label{subsec:incomplete_skymodel_cal}
In this simulation, we assess the impact of unmodeled diffuse Galactic emission during DI-gain calibration. We therefore forward-simulated a PSDE sky model to obtain visibilities but DI-gain calibrated the data with a PSO sky model. This produces a mismatch between the forward simulation and the calibration process. We then evaluated the systematic power spectrum bias, $b^{\rm sys}$, resulting from calibration errors due to the omission of diffuse Galactic emission.

To achieve this, we generated ten power spectra from DI-calibrated visibilities ($V_{ij}^{\rm DI-cal}$), where each power spectrum corresponds to a different Gaussian realization of the diffuse foregrounds. The point-source sky and noise realization were held constant to isolate the calibration errors caused by unmodeled diffuse Galactic emission from other sources of variation. This setup ensures that we can attribute any residual power spectrum bias to the errors introduced by the mismatched sky model with a statistical sample variance of 10\%.

However, to accurately disentangle the bias caused by calibration errors from the effects of thermal noise variations, we performed an additional simulation. In this extended setup, we introduced variations in both Gaussian foregrounds and thermal noise. This allows us to determine how much of the observed bias arises from thermal noise fluctuations versus calibration errors due to the omission of Galactic diffuse emission during DI-gain calibration.

The systematic power spectrum bias for these two setups is again estimated by calculating the mean residuals from the fiducial true power spectrum, according to the Equation \ref{eq:bsys}. $P_i$ is a single estimated power spectrum calculated from miscalibrated visibilities. The two setups are:

\begin{itemize}
    \item Setup 2: $\left< \Delta P_{D_i}^{\rm DI-cal} \right>$ is the expectation value of power spectrum residuals, when DI-gain calibration is performed with a mismatched sky model $\left< \Delta P_{D_i}^{\rm DI-cal} \right>$ under different realizations of diffuse foregrounds $D_i$. The fiducial power spectrum $\left< P_{\rm true} \right> = \left< P_{\rm P} + P_{D_i} + P_{\rm N} \right>$ is the ensemble average of power spectra generated from true visibilities $V_{ij}^{\rm true}$ under variation of diffuse foregrounds.
    \item Setup 3: $\left< \Delta P_{D_i, N_i}^{\rm DI-cal} \right>$ is the expectation value of power spectrum residuals under different realizations of both the diffuse foregrounds and thermal noise, when DI-gain calibration is performed with a mismatched sky model. The fiducial power spectrum $\left< P_{\rm true} \right> = \left< P_{\rm P} + P_{D_i} + P_{N_i} \right>$ is the ensemble average of power spectra generated from true visibilities with different realizations of diffuse foregrounds and thermal noise.
\end{itemize}

The results of these three setups are shown in Figure \ref{fig:delta_p}. The systematic bias, as defined in Equation \ref{eq:bsys}, is expressed as a fraction of the thermal noise power for 4 hours of observations with LOFAR-HBA observations. This represents the average 2D cylindrical power spectrum of Stokes V when the thermal noise realization is varied in visibility space. The systematic bias ranges from $10^{-3}$ (blue regions) to $10^3$ (red regions) in the graphs. Blue regions represent power spectrum bins in which the systematic error due to DI-gain calibration errors is subdominant to the thermal noise after 4 hours of observations, whereas red regions represent bins in which the systematic error dominates over the thermal noise. Calibrating with a mismatched sky model that is omitting the diffuse Galactic emission (Setup 2 and 3), increases the systematic bias by approximately one order of magnitude (see also Figure \ref{fig:bsys_sigma}) compared to matched PSO sky model calibration (Setup 1), although the systematic bias is mostly confined to $k_{\parallel} < 0.1\, h \mathrm{Mpc}^{-1}$. Residual foregrounds, which dominate in regions $k_{\parallel} < 0.2 \, h\mathrm{Mpc}^{-1}$, can be removed using Gaussian process regression \citep{2018:mertens, 2020:mertens, 2024:mertens}, provided they are coherent between observations and the excess power exhibits a frequency-frequency coherency scale similar to that of the foregrounds. The coherency between pairs of 4-hour observations will be analyzed in the next section.

\subsection{Coherence of DI-gain calibration errors}
To better understand whether the excess power from DI-gain calibration errors is correlated between nightly observations, we conduct two tests. First, we examine whether the systematic bias integrates down or not as integration time increases. Second, we evaluate the cross-coherency between 4-hour observations taken at different local sidereal times.

\subsubsection{Combining data sets}
\label{subsec:combine_datasets}
In this section, we present the power spectra obtained by combining three observations taken at different local sidereal times (LST), totaling 12 hours of simulated data. Unlike repeated observations of the same LST with different noise or foreground realizations (Section \ref{subsec:complete_skymodel_cal} and \ref{subsec:incomplete_skymodel_cal}), these three observations are spaced 4 hours in LST apart to better reflect realistic observing conditions for a symmetric array such as LOFAR. It is necessary to combine observations to reduce the thermal noise level and to determine if the systematic bias due to DI-gain calibration errors remains or is reduced when combining observations from different LST intervals. It is expected that a total of about 1000 hours of LOFAR-HBA observation on a deep field is required for a statistical detection of the 21 cm signal from the EoR \citep{2020:mertens}. These LST intervals account for changes in the PSF over time, with the specifications of the simulated datasets listed in Table \ref{tab:simulated_dataset}. 

We used inverse variance weighting to optimally average the datasets in visibility space, ensuring an accurate combination of data from different observations for power spectrum estimation. Observation-to-observation variations in noise levels were addressed by using Stokes-V sub-band difference noise estimates, which were combined with weights based on the ($u,v$) density of the gridded visibilities. The methodology for estimating these weights is described in detail in Section 3.2.3 of \citet{2020:mertens}.

To determine the systematic bias, we created a stack of observations both before and after DI-gain calibration. Since our simulations are only 4 hours long, some $uv$-cells contain fewer visibilities. To mitigate this issue, we filtered visibilities with a minimum weight of 100, resulting in gaps in the $uv$-coverage and causing thermal noise to integrate down more slowly in certain regions of power spectrum space when combining observations from different LSTs. The pre-calibration stack serves as our fiducial power spectrum, $P_{\rm true}$, while the post-calibration stack represents the measured power spectrum with the DI-gain calibration errors embedded. The systematic bias for stacks of 1, 2, and 3 observations with different LST ranges, expressed as a fraction of the thermal noise, is shown in Figure \ref{fig:stack_complete} with matched PSO sky model calibration, and in Figure \ref{fig:stack_incomplete} with mismatched sky model calibration, which omits the diffuse Galactic emission component. These are detailed in the Appendix \ref{sec:appendix_a}. From these, we observe that the systematic bias, for both the matched and mismatched sky model calibration cases, does not clearly decrease with increasing observation time. This suggests that DI-gain calibration errors in the foreground dominated region $k_{\parallel}$-bins $< 0.2 \, h\mathrm{Mpc}^{-1}$ are coherent, while thermal noise, which dominates in $k_{\parallel}$-bins $> 0.2 \, h\mathrm{Mpc}^{-1}$, is incoherent and decreases with longer integration times.

\begin{figure*}
    \centering
    \includegraphics[trim={5cm 0cm 0cm 0cm}, clip, scale=0.23]{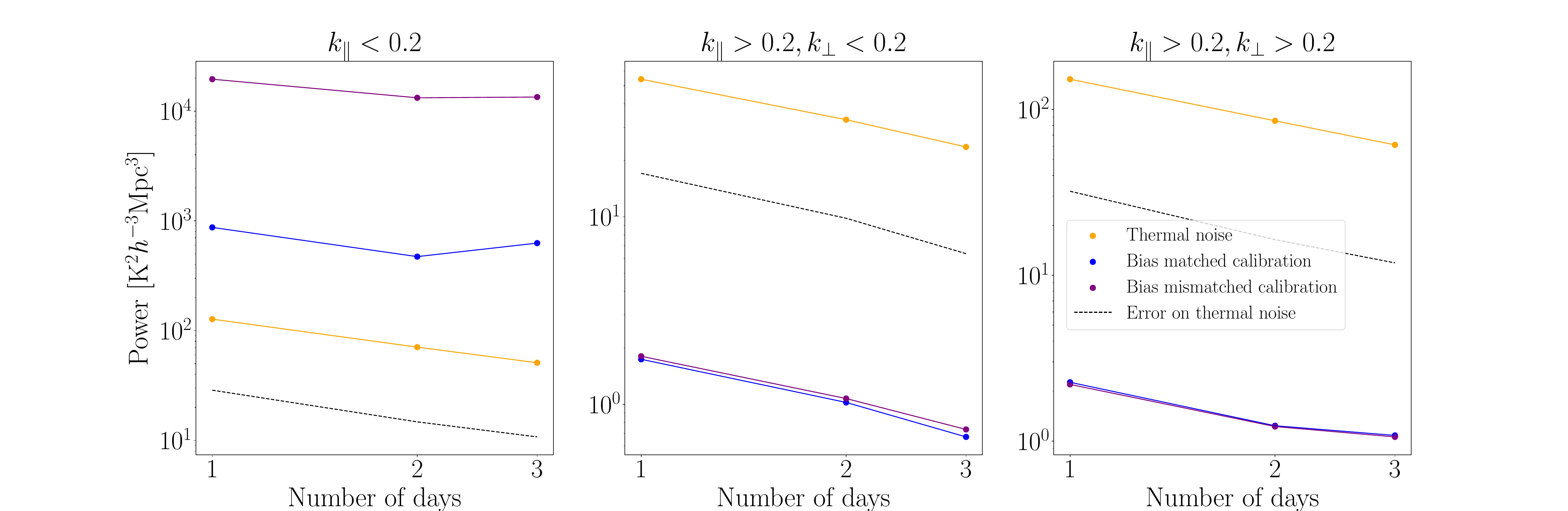}A
    \caption{The average systematic bias, the expected thermal noise and the error on the thermal noise as a function of observation time for three different regions in the 2D power spectrum space: The foreground dominated region in panel 1 ($k_{\parallel}$-bins $< 0.2 \, h\mathrm{Mpc}^{-1}$), the short baseline region, where diffuse Galactic emission is prominent in panel 2 ($k_{\parallel}$-bins $> 0.2 \, h\mathrm{Mpc}^{-1}$ and $k_{\perp}$-bins $< 0.1 \, h\mathrm{Mpc}^{-1}$) and the noise dominated region in panel 3 ($k_{\parallel}$-bins $> 0.2 \, h\mathrm{Mpc}^{-1}$ and $k_{\perp}$-bins $> 0.1 \, h\mathrm{Mpc}^{-1}$). In orange we show the expected thermal noise, in dashed black the error on the thermal noise as a function of integration time and in blue and purple the systematic bias for matched PSO and mismatched sky model calibration, respectively.}
    \label{fig:bsys_sigma}
\end{figure*}

To better understand the magnitude of the bias compared to thermal noise and to determine whether it integrates down, we divided the $(k_{\perp}, k_{\parallel})$-space into three regions: the foreground dominated region for $k_{\parallel}$-bins $< 0.2 \, h\mathrm{Mpc}^{-1}$, and two EoR-window regions distinguishing between the shorter baselines ($|\mathbf{u}| < 100$; roughly the LOFAR 'superterp' region) with $k_{\parallel}$-bins $> 0.2 \, h\mathrm{Mpc}^{-1}$ and $k_{\perp}$-bins $< 0.1 \, h\mathrm{Mpc}^{-1}$ and the longer baselines with $k_{\parallel}$-bins $> 0.2 \, h\mathrm{Mpc}^{-1}$ and $k_{\perp}$-bins $> 0.1 \, h\mathrm{Mpc}^{-1}$. The region where $|\mathbf{u}| < 100$, corresponds to prominent diffuse Galactic emission; however, for $|\mathbf{u}| > 100$, noise becomes dominant. We averaged the systematic bias (as defined in Equation \ref{eq:bsys}) and the thermal noise over power spectrum bins for these three regions and plotted them as a function of number of observations, which corresponds to observation time, on a log-log plot in Figure \ref{fig:bsys_sigma}.

In the left panel of Figure \ref{fig:bsys_sigma}, which represents the foreground dominated region for $k_{\parallel}$-bins $< 0.2 \, h\mathrm{Mpc}^{-1}$, we clearly observe that the thermal noise decreases with integration time, while the systematic bias is above thermal noise and remains approximately constant. Even though the bias is much higher than thermal noise in the foreground dominated region, GPR is able to remove these signals if they are coherent in frequency and observation time. The frequency coherence is ensured by the smooth gain solutions, ensuring no rapidly varying frequency structures.

In the middle panel, we compare the systematic bias and thermal noise for $|\mathbf{u}| < 100$, corresponding to $k_{\parallel}$-bins $> 0.2 \, h\mathrm{Mpc}^{-1}$ and $k_{\perp}$-bins $< 0.1 \, h\mathrm{Mpc}^{-1}$. Here, we find that the systematic bias is $1-2$ orders of magnitude lower than the thermal noise and $\sim$ one order of magnitude below the error of thermal noise and it is evident that the thermal noise decreases with a similar slope as the bias as a function of integration time (days). There is a slight difference in the bias between calibrating with a matched sky model versus a mismatched one, although these differences are minimal. Based on these these observations we did a linear fit of the bias, thermal noise power and error on the thermal noise versus hours of integration. The slopes were found to be very similar, indicating that the average power spectrum residuals integrate down in a manner similar to thermal noise. The same conclusion applies for the noise dominated region ($k_{\parallel}$-bins $> 0.2 \, h\mathrm{Mpc}^{-1}$ and $k_{\perp}$-bins $> 0.1 \, h\mathrm{Mpc}^{-1}$): the slopes of the thermal noise, error on the thermal noise and bias power are almost identical, meaning the average power spectrum residuals in this region also integrate down like thermal noise. The error on the thermal noise is due to sampling variance, which depends on the number of $uv$-cells averaged into a power spectrum bin. Due to different visibility flagging for different LST ranges, the slope of the error on the thermal noise differs between stacking two observations versus three.

To further investigate whether the DI-calibration bias continues to decrease with observation time or eventually plateaus, we also conducted a set of simulations using repeated 4-hour observations at the same LST. In this case, both the sky and the PSF remain fixed, and only the thermal noise is different between the observations. Under these conditions, we observe that the systematic bias still decreases, but with a shallower slope than in the case where observations span different LSTs. This implies that time-varying PSFs, resulting from observing at different LSTs, introduce incoherence in the calibration errors due to mode-mixing, thereby helping the bias integrate down more effectively. To quantify the rate of decrease, we performed a linear fit of the bias and thermal noise power as a function of integration time (in hours). We find that the bias becomes equal to the thermal noise level after approximately 90 hours of integration for the matched PSO calibration and after about 40 hours for the mismatched sky model calibration. This illustrates how calibration errors accumulate differently depending on calibration sky model completeness, and how PSF variation can help mitigate their coherence over time. Extrapolating from these results, after stacking 4-hour observations at the same LST, the bias in the short baseline region reaches the thermal noise level after roughly 10 times the length of a single observation. By analogy, for 12-hour observations at the same LST, we expect the bias to decrease at a similar rate per observation, reaching the thermal noise level after roughly 120 hours of total integration time. This extrapolation assumes the bias behavior from 4-hour blocks scales linearly or comparably for longer periods. While plausible, this assumption should be validated with further simulations or data which is beyond the scope of this work. 

We conclude this section that even with thermal noise variations and matched PSO sky model calibration a systematic bias is introduced. Thermal noise variations lead to small gain errors, which, when applied to the bright point source model, introduce a power spectrum bias. However, these errors are only significant at $k_{\parallel}$-bins $< 0.2 \, h\mathrm{Mpc}^{-1}$ and could be mitigated by GPR. The same applies to mismatched sky model calibration, where omitting diffuse Galactic emission from the sky model leads to a bias approximately one magnitude higher in the same region.

\subsubsection{Coherence across observations}

For Gaussian process regression (GPR) to effectively remove excess power within the foreground-dominated region, the signal must exhibit a frequency-frequency coherence scale similar to that of the foregrounds and must be coherent across observations \citep{2020:mertens}. We determine the correlation between all pairs of simulated observations by computing the cylindrically-averaged cross-coherence, defined as:

\begin{equation}
\label{eq:coherence}
C_{1,2}(k_{\perp}, k_{\parallel}) = \frac{\langle | \tilde{T}_1^*(\mathbf{k}) \tilde{T}_2(\mathbf{k}) |\rangle}{\sqrt{\langle | \tilde{T}_1(\mathbf{k})|^2\rangle \langle | \tilde{T}_2(\mathbf{k})|^2 \rangle }}.
\end{equation}

\begin{figure*}[h!]
    \centering
        \includegraphics[width=0.95\textwidth, trim={2cm 3cm 1cm 2cm}, clip, scale=0.25]{202507_avg_coherence_complete_skymodel.png} \\
        \includegraphics[width=0.95\textwidth, trim={2cm 3cm 1cm 2cm}, clip, scale=0.25]{20250306_avg_coherence_incomplete_skymodel.png}
        \caption{Cross-coherence corner-plot between three different simulated observations covering different local sidereal time (LST) ranges. Top panel: Average cross-coherence when DI-gain calibrating the visibilities with a matched PSO sky model. Bottom panel: Average cross-coherence when DI-gain calibrating the visibilities with an mismatched sky model, which is omitting the diffuse Galactic emission. 
        Three regions of the $(k_{\perp}, k_{\parallel})$-space are analyzed: The foreground region $k_{\parallel}< 0.2 \, h\mathrm{Mpc}^{-1}$ (left panel), and two EoR-windows for $k_{\parallel} > 0.2 \, h\mathrm{Mpc}^{-1}$ and baselines $< 100 \, \lambda$ (middle panel), and baselines $> 100 \, \lambda$ (right panel). There is on average strong cross-coherence between observations in the foreground region (left panel), whereas weak coherence in the EoR-regions (middle and right panel). There are only minor differences in the average cross-coherence in the EoR-window for baselines $< 100 \, \lambda$, where diffuse Galactic emission is prominent, between calibrating with a matched PSO sky model versus a mismatched one.}
        \label{fig:coherence_cornerplot}
\end{figure*}

The cross-coherence is a normalized metric ranging from -1 to 1, where -1 indicates strong anti-correlation, 1 indicates strong correlation, and 0 signifies no correlation. If the bias is only partially coherent in either of the two EoR-window regions, it not only averages down more slowly than incoherent thermal noise (as demonstrated in the previous section), but also introduces a bias in the 21 cm signal power spectrum that increases with longer integration times and cannot be mitigated by standard GPR. This is because, in the current NCP pipeline, GPR is applied after combining data from multiple nights, which limits its ability to mitigate a partially coherent bias. We note that applying GPR on a per-night basis may alleviate this and is currently under investigation.

To test for cross-coherence across observations, we use the three simulated observations taken at different LST, and compute the cross-coherence between all pairs of observations in the case of matched PSO sky model calibration and mismatched sky model calibration. We show the average cross-coherence over all observation pairs in Figure \ref{fig:avg_cross_coh} of the Appendix \ref{sec:appendix_b}. When differencing the average cross-coherence from simulated observations DI-gain calibrated with a mismatched sky model from the ones calibrated with a matched PSO sky model, we see up to 10\% higher coherence in the average cross-coherence obtained from power spectra calibrated with a mismatched sky model, omitting the diffuse Galactic emission. The difference is specifically noticeable on shorter baselines ($|\mathbf{u}| < 100$ and $k_{\perp} < 0.1 \, h\mathrm{Mpc}^{-1}$), pointing to partial coherence due to the missing diffuse Galactic emission, which is distinct from incoherent thermal noise at baselines $|\mathbf{u}| > 100$.

To further compress the information from the cross-coherence between all pairs of observations, we again divide the $(k_{\perp}, k_{\parallel})$-space in a foreground dominated region ($k_{\parallel}< 0.2 \, h\mathrm{Mpc}^{-1}$) and the two EoR-window regions separated by baseline length ($k_{\parallel} > 0.2 \, h\mathrm{Mpc}^{-1}$ and $|\mathbf{u}| < 100$ or $|\mathbf{u}| > 100$) and average over the bins in these three regions. A corner-plot of the correlations between observations for each of the three different regions is presented in Figure \ref{fig:coherence_cornerplot} in the top panel for matched PSO sky model calibration and in the bottom panel for mismatched sky model calibration. In the foreground region (left panels), the signal is predominantly coherent, whereas the EoR-windows exhibit incoherence on average across observations. The short baseline region (middle panel) displays greater coherence on average compared to the noise-dominated region (right panel), with only minor differences observed between calibrating with a matched sky model and a mismatched one.

\section{Summary and conclusions}
\label{sec:summary}
We developed a realistic forward simulation pipeline called \texttt{Simple} (SIMulation Pipeline for LOFAR-EoR) using \texttt{Nextflow}, a workflow management tool, to study the impact of diffuse Galactic emission on direction-independent (DI) gain calibration for LOFAR-EoR observations. DI-gain calibration is performed with the non-linear optimization method named \texttt{SAGECAL-CO}, which is also used for real LOFAR-EoR observations.

Our approach mirrors the two-step DI-gain calibration scheme employed in LOFAR-EoR data processing. First, smooth spectral and rapidly time-varying gain variations are corrected, followed by a band-pass calibration step to address fast band-pass and slowly time-varying variations. The gain solutions from both DI-steps are combined and applied directly to the simulated visibilities, embedding any DI-gain calibration errors in the data and making them challenging to mitigate in subsequent data processing tasks.

Baselines ranging from 50 to 5000 $\lambda$ are used during DI-gain calibration. Diffuse Galactic emission dominates over confusion noise and could therefore represent the dominant component of the missing sky model on baselines between 50 and 250 $\lambda$. This baseline range also corresponds to the range used for EoR 21 cm signal power spectrum analysis. Data are forward-simulated and then calibrated under two scenarios.
Scenario one is the ground truth case, obtained by forward-simulating and calibrating with a matched PSO sky model. Scenario two is obtained by forward-simulating a PSDE sky model and calibrating with a PSO sky model, thereby omitting diffuse Galactic emission during DI-gain calibration. This creates a mismatch between forward-simulation and calibration of the data.
The two scenarios enable us to evaluate the impact of unmodeled diffuse Galactic emission relative to a control simulation with a sky model that is complete and accurate.

 We analyzed gain solutions, images, and power spectra for the two calibration scenarios, yielding the following results:

\begin{enumerate}
    \item Gain solution variability:
    The standard deviation of the histogram of the diagonal gain solutions increases by a factor of $\sim$2 when calibrating with a mismatched sky model. This indicates that the diffuse emission's power is redistributed into the gain solutions, making them noisier. Baselines involving the Superterp and nearby core stations (most sensitive to diffuse Galactic emission) show an increase in gain solution variability from 0.5\% to 1\% with a mismatched sky model calibration.
    \item Residual diffuse emission: Residual diffuse Galactic emission appears in image space for baselines between 50 and $250~\lambda$ when comparing images before and after DI-gain calibration. The RMS of the residual emission is $\sim$ \SI{6e-3}{\jy}/PSF, roughly an order of magnitude smaller than the original modeled diffuse Galactic emission ($\sim$\SI{1e-1}{\jy}/PSF). Almost no residual emission is observed in the difference images when calibrating with a matched PSO sky model (see Figure \ref{fig:diff_img_ps}).
    \item Power spectrum analysis:
    The systematic bias in power spectrum space, relative to thermal noise, is approximately an order of magnitude higher in the foreground region ($k_{\parallel}$ $<0.2 \, h\mathrm{Mpc}^{-1}$) for mismatched sky model calibration compared to matched PSO sky model calibration. In the foreground region, the systematic bias does not reduce with integration time like thermal noise. However, if this bias is coherent across observations (as shown in Figures \ref{fig:avg_cross_coh} and \ref{fig:coherence_cornerplot}), Gaussian Process Regression (GPR) can likely effectively remove it \citep{2018:mertens, 2020:mertens, 2024:mertens}.
    \item Integration time requirements:
    In the EoR-window, for short ($|\mathbf{u}| < 100$) and long baselines ($|\mathbf{u}| > 100$), the bias is $1-2$ orders of magnitude below the thermal noise and integrates down at the same rate as thermal noise for both DI-gain calibration scenarios (matched/mismatched sky model) as a function of observing time. When stacking observations of the same LST range with a fixed PSF, the bias decreases more slowly than thermal noise. Observations with time-varying PSFs cause incoherent DI-gain errors that reduce the bias more efficiently, but after 12 hours, PSF changes no longer help. Based on our analysis in section \ref{subsec:combine_datasets}, we estimate that the DI-gain calibration bias, due to the missing diffuse Galactic component, is expected to approach the level of thermal noise after approximately 120 hours of total integration time; however, this extrapolation remains subject to validation with additional simulations.
    \item Coherence analysis: The cross-coherence analysis shows no significant difference in the foreground region ($k_{\parallel}$-bins $< 0.2 \, h\mathrm{Mpc}^{-1}$) between matched and mismatched sky model calibration, indicating that excess power can likely be effectively removed with GPR \citep{2018:mertens}. In the EoR-window for short baselines ($|\mathbf{u}| < 100$), coherence slightly increases with calibration errors from the missing diffuse Galactic emission component compared to matched PSO sky model calibration in some power spectrum bins (see \ref{fig:avg_cross_coh}). However, we believe these changes in coherence do not significantly affect the performance of GPR in foreground removal.
\end{enumerate}

Our results indicate that the absence of diffuse Galactic emission in the sky model during calibration is unlikely to be responsible for the excess variance observed in LOFAR’s current \SI{21}{\centi\m} signal power spectrum upper limits. In real observations, we model the diffuse Galactic emission as part of the Gaussian Process Regression-based foreground removal step to subtract it from the data. However, this conclusion is based on a simulation that assumed perfect knowledge of the beam. Specifically, the same beam model used during forward simulation was also applied during data analysis, ensuring consistency between the simulated visibilities and the calibration/imaging steps. This approach eliminates any discrepancies arising from mismatched beam assumptions. However, in real-world observations, beam errors can occur, such as tiles being turned off during specific observations. These errors could alter the influence of diffuse Galactic emission on DI-gain calibration. Addressing the impact of such beam errors will be the focus of a follow-up study.


%


%

\begin{acknowledgements}
      CH, LVEK, SAB, KC, SG and SM acknowledge the financial support from the European Research Council (ERC) under the European Union’s Horizon 2020 research and innovation programme (Grant agreement No. 884760, ”CoDEX”). EC would like to acknowledge the support from the Centre for Data Science and Systems Complexity (DSSC), Faculty of Science and Engineering at the University of Groningen, and from the Ministry of Universities and Research (MUR) through the PRIN project 'Optimal inference from radio images of the epoch of reionization'. FGM acknowledges the financial support of the PSL Fellowship Programme.
\end{acknowledgements}

%
%


\begin{thebibliography}{68}
\expandafter\ifx\csname natexlab\endcsname\relax\def\natexlab#1{#1}\fi

\bibitem[{Atek {et~al.}(2022)Atek, Shuntov, Furtak, Richard, Kneib, Mahler, Zitrin, McCracken, Charlot, Chevallard, \& Chemerynska}]{2022:atek}
Atek, H., Shuntov, M., Furtak, L.~J., {et~al.} 2022, \mnras, 519, 1201

\bibitem[{Barry {et~al.}(2016)Barry, Hazelton, Sullivan, Morales, \& Pober}]{2016:barry}
Barry, N., Hazelton, B., Sullivan, I., Morales, M.~F., \& Pober, J.~C. 2016, \mnras, 461, 3135

\bibitem[{{Beardsley} {et~al.}(2016){Beardsley}, {Hazelton}, {Sullivan}, {Carroll}, {Barry}, {Rahimi}, {Pindor}, {Trott}, {Line}, {Jacobs}, {Morales}, {Pober}, {Bernardi}, {Bowman}, {Busch}, {Briggs}, {Cappallo}, {Corey}, {de Oliveira-Costa}, {Dillon}, {Emrich}, {Ewall-Wice}, {Feng}, {Gaensler}, {Goeke}, {Greenhill}, {Hewitt}, {Hurley-Walker}, {Johnston-Hollitt}, {Kaplan}, {Kasper}, {Kim}, {Kratzenberg}, {Lenc}, {Loeb}, {Lonsdale}, {Lynch}, {McKinley}, {McWhirter}, {Mitchell}, {Morgan}, {Neben}, {Thyagarajan}, {Oberoi}, {Offringa}, {Ord}, {Paul}, {Prabu}, {Procopio}, {Riding}, {Rogers}, {Roshi}, {Udaya Shankar}, {Sethi}, {Srivani}, {Subrahmanyan}, {Tegmark}, {Tingay}, {Waterson}, {Wayth}, {Webster}, {Whitney}, {Williams}, {Williams}, {Wu}, \& {Wyithe}}]{2016:beardsley}
{Beardsley}, A.~P., {Hazelton}, B.~J., {Sullivan}, I.~S., {et~al.} 2016, \apj, 833, 102

\bibitem[{{Becker} {et~al.}(2001){Becker}, {Fan}, {White}, {Strauss}, {Narayanan}, {Lupton}, {Gunn}, {Annis}, {Bahcall}, {Brinkmann}, {Connolly}, {Csabai}, {Czarapata}, {Doi}, {Heckman}, {Hennessy}, {Ivezi{\'c}}, {Knapp}, {Lamb}, {McKay}, {Munn}, {Nash}, {Nichol}, {Pier}, {Richards}, {Schneider}, {Stoughton}, {Szalay}, {Thakar}, \& {York}}]{2001:becker}
{Becker}, R.~H., {Fan}, X., {White}, R.~L., {et~al.} 2001, \aj, 122, 2850

\bibitem[{{Bernardi} {et~al.}(2009{\natexlab{a}}){Bernardi}, {de Bruyn}, {Brentjens}, {Ciardi}, {Harker}, {Jeli{\'c}}, {Koopmans}, {Labropoulos}, {Offringa}, {Pandey}, {Schaye}, {Thomas}, {Yatawatta}, \& {Zaroubi}}]{2009:bernadi}
{Bernardi}, G., {de Bruyn}, A.~G., {Brentjens}, M.~A., {et~al.} 2009{\natexlab{a}}, \aap, 500, 965

\bibitem[{{Bernardi} {et~al.}(2009{\natexlab{b}}){Bernardi}, {de Bruyn}, {Brentjens}, {Ciardi}, {Harker}, {Jeli{\'c}}, {Koopmans}, {Labropoulos}, {Offringa}, {Pandey}, {Schaye}, {Thomas}, {Yatawatta}, \& {Zaroubi}}]{2009:bernadi2}
{Bernardi}, G., {de Bruyn}, A.~G., {Brentjens}, M.~A., {et~al.} 2009{\natexlab{b}}, \aap, 500, 965

\bibitem[{{Bowman} {et~al.}(2018){Bowman}, {Rogers}, {Monsalve}, {Mozdzen}, \& {Mahesh}}]{2018:bowman}
{Bowman}, J.~D., {Rogers}, A. E.~E., {Monsalve}, R.~A., {Mozdzen}, T.~J., \& {Mahesh}, N. 2018, \nat, 555, 67

\bibitem[{Boyd {et~al.}(2011)Boyd, Parikh, Chu, Peleato, \& Eckstein}]{2011:boyd}
Boyd, S., Parikh, N., Chu, E., Peleato, B., \& Eckstein, J. 2011, Foundations and Trends® in Machine Learning, 3, 1

\bibitem[{{Brackenhoff} {et~al.}(2024){Brackenhoff}, {Mevius}, {Koopmans}, {Offringa}, {Ceccotti}, {Chege}, {Gehlot}, {Ghosh}, {H{\"o}fer}, {Mertens}, {Munshi}, \& {Zaroubi}}]{2024:Brackenhoff}
{Brackenhoff}, S.~A., {Mevius}, M., {Koopmans}, L.~V.~E., {et~al.} 2024, \mnras, 533, 632

\bibitem[{{Byrne} {et~al.}(2022){Byrne}, {Morales}, {Hazelton}, {Sullivan}, {Barry}, {Lynch}, {Line}, \& {Jacobs}}]{2022:bryne}
{Byrne}, R., {Morales}, M.~F., {Hazelton}, B., {et~al.} 2022, \mnras, 510, 2011

\bibitem[{{Ceccotti} {et~al.}(2023){Ceccotti}, {Offringa}, {Koopmans}, {Timmerman}, {Brackenhoff}, {Gehlot}, {Mertens}, {Munshi}, {Pandey}, {van Weeren}, {Yatawatta}, \& {Zaroubi}}]{2023:ceccotti}
{Ceccotti}, E., {Offringa}, A.~R., {Koopmans}, L.~V.~E., {et~al.} 2023, \mnras, 525, 3946

\bibitem[{{de Oliveira-Costa} {et~al.}(2008){de Oliveira-Costa}, {Tegmark}, {Gaensler}, {Jonas}, {Landecker}, \& {Reich}}]{2008:oliveira-costa}
{de Oliveira-Costa}, A., {Tegmark}, M., {Gaensler}, B.~M., {et~al.} 2008, \mnras, 388, 247

\bibitem[{{DeBoer} {et~al.}(2017){DeBoer}, {Parsons}, {Aguirre}, {Alexander}, {Ali}, {Beardsley}, {Bernardi}, {Bowman}, {Bradley}, {Carilli}, {Cheng}, {de Lera Acedo}, {Dillon}, {Ewall-Wice}, {Fadana}, {Fagnoni}, {Fritz}, {Furlanetto}, {Glendenning}, {Greig}, {Grobbelaar}, {Hazelton}, {Hewitt}, {Hickish}, {Jacobs}, {Julius}, {Kariseb}, {Kohn}, {Lekalake}, {Liu}, {Loots}, {MacMahon}, {Malan}, {Malgas}, {Maree}, {Martinot}, {Mathison}, {Matsetela}, {Mesinger}, {Morales}, {Neben}, {Patra}, {Pieterse}, {Pober}, {Razavi-Ghods}, {Ringuette}, {Robnett}, {Rosie}, {Sell}, {Smith}, {Syce}, {Tegmark}, {Thyagarajan}, {Williams}, \& {Zheng}}]{2017:deboer}
{DeBoer}, D.~R., {Parsons}, A.~R., {Aguirre}, J.~E., {et~al.} 2017, \pasp, 129, 045001

\bibitem[{Donnan {et~al.}(2022)Donnan, McLeod, Dunlop, McLure, Carnall, Begley, Cullen, Hamadouche, Bowler, Magee, McCracken, Milvang-Jensen, Moneti, \& Targett}]{2022:donnan}
Donnan, C.~T., McLeod, D.~J., Dunlop, J.~S., {et~al.} 2022, \mnras, 518, 6011

\bibitem[{{Dowell} {et~al.}(2017){Dowell}, {Taylor}, {Schinzel}, {Kassim}, \& {Stovall}}]{2017:dowell}
{Dowell}, J., {Taylor}, G.~B., {Schinzel}, F.~K., {Kassim}, N.~E., \& {Stovall}, K. 2017, \mnras, 469, 4537

\bibitem[{{Eastwood} {et~al.}(2018){Eastwood}, {Anderson}, {Monroe}, {Hallinan}, {Barsdell}, {Bourke}, {Clark}, {Ellingson}, {Dowell}, {Garsden}, {Greenhill}, {Hartman}, {Kocz}, {Lazio}, {Price}, {Schinzel}, {Taylor}, {Vedantham}, {Wang}, \& {Woody}}]{2018:eastwood}
{Eastwood}, M.~W., {Anderson}, M.~M., {Monroe}, R.~M., {et~al.} 2018, \aj, 156, 32

\bibitem[{{Edler} {et~al.}(2021){Edler}, {de Gasperin}, \& {Rafferty}}]{2021:edler}
{Edler}, H.~W., {de Gasperin}, F., \& {Rafferty}, D. 2021, \aap, 652, A37

\bibitem[{Ewall-Wice {et~al.}(2017)Ewall-Wice, Dillon, Liu, \& Hewitt}]{2017:ewallwice}
Ewall-Wice, A., Dillon, J.~S., Liu, A., \& Hewitt, J. 2017, \mnras, 470, 1849

\bibitem[{{Finkelstein} {et~al.}(2024){Finkelstein}, {Leung}, {Bagley}, {Dickinson}, {Ferguson}, {Papovich}, {Akins}, {Arrabal Haro}, {Dav{\'e}}, {Dekel}, {Kartaltepe}, {Kocevski}, {Koekemoer}, {Pirzkal}, {Somerville}, {Yung}, {Amor{\'\i}n}, {Backhaus}, {Behroozi}, {Bisigello}, {Bromm}, {Casey}, {Ch{\'a}vez Ortiz}, {Cheng}, {Chworowsky}, {Cleri}, {Cooper}, {Davis}, {de la Vega}, {Elbaz}, {Franco}, {Fontana}, {Fujimoto}, {Giavalisco}, {Grogin}, {Holwerda}, {Huertas-Company}, {Hirschmann}, {Iyer}, {Jogee}, {Jung}, {Larson}, {Lucas}, {Mobasher}, {Morales}, {Morley}, {Mukherjee}, {P{\'e}rez-Gonz{\'a}lez}, {Ravindranath}, {Rodighiero}, {Rowland}, {Tacchella}, {Taylor}, {Trump}, \& {Wilkins}}]{2024:finkelstein}
{Finkelstein}, S.~L., {Leung}, G. C.~K., {Bagley}, M.~B., {et~al.} 2024, \apjl, 969, L2

\bibitem[{{Gan} {et~al.}(2022){Gan}, {Koopmans}, {Mertens}, {Mevius}, {Offringa}, {Ciardi}, {Gehlot}, {Ghara}, {Ghosh}, {Giri}, {Iliev}, {Mellema}, {Pandey}, \& {Zaroubi}}]{2022:gan}
{Gan}, H., {Koopmans}, L.~V.~E., {Mertens}, F.~G., {et~al.} 2022, \aap, 663, A9

\bibitem[{{Gehlot} {et~al.}(2022){Gehlot}, {Koopmans}, {Offringa}, {Gan}, {Ghara}, {Giri}, {Kuiack}, {Mertens}, {Mevius}, {Mondal}, {Pandey}, {Shulevski}, {Wijers}, \& {Yatawatta}}]{2022:gehlot}
{Gehlot}, B.~K., {Koopmans}, L.~V.~E., {Offringa}, A.~R., {et~al.} 2022, \aap, 662, A97

\bibitem[{{Grobler} {et~al.}(2014){Grobler}, {Nunhokee}, {Smirnov}, {van Zyl}, \& {de Bruyn}}]{2014:Grobler}
{Grobler}, T.~L., {Nunhokee}, C.~D., {Smirnov}, O.~M., {van Zyl}, A.~J., \& {de Bruyn}, A.~G. 2014, \mnras, 439, 4030

\bibitem[{{Gupta} {et~al.}(2017){Gupta}, {Ajithkumar}, {Kale}, {Nayak}, {Sabhapathy}, {Sureshkumar}, {Swami}, {Chengalur}, {Ghosh}, {Ishwara-Chandra}, {Joshi}, {Kanekar}, {Lal}, \& {Roy}}]{2017:gupta}
{Gupta}, Y., {Ajithkumar}, B., {Kale}, H.~S., {et~al.} 2017, Current Science, 113, 707

\bibitem[{{Hamaker} {et~al.}(1996){Hamaker}, {Bregman}, \& {Sault}}]{1996:hamaker}
{Hamaker}, J.~P., {Bregman}, J.~D., \& {Sault}, R.~J. 1996, \aaps, 117, 137

\bibitem[{{Harikane} {et~al.}(2023){Harikane}, {Zhang}, {Nakajima}, {Ouchi}, {Isobe}, {Ono}, {Hatano}, {Xu}, \& {Umeda}}]{2023:harikane}
{Harikane}, Y., {Zhang}, Y., {Nakajima}, K., {et~al.} 2023, \apj, 959, 39

\bibitem[{{Haslam} {et~al.}(1981){Haslam}, {Klein}, {Salter}, {Stoffel}, {Wilson}, {Cleary}, {Cooke}, \& {Thomasson}}]{1981:haslamI}
{Haslam}, C.~G.~T., {Klein}, U., {Salter}, C.~J., {et~al.} 1981, \aap, 100, 209

\bibitem[{{Haslam} {et~al.}(1982){Haslam}, {Salter}, {Stoffel}, \& {Wilson}}]{1982:HaslamII}
{Haslam}, C.~G.~T., {Salter}, C.~J., {Stoffel}, H., \& {Wilson}, W.~E. 1982, \aaps, 47, 1

\bibitem[{{HERA Collaboration} {et~al.}(2023){HERA Collaboration}, {Abdurashidova}, {Adams}, {Aguirre}, {Alexander}, {Ali}, {Baartman}, {Balfour}, {Barkana}, {Beardsley}, {Bernardi}, {Billings}, {Bowman}, {Bradley}, {Breitman}, {Bull}, {Burba}, {Carey}, {Carilli}, {Cheng}, {Choudhuri}, {DeBoer}, {de Lera Acedo}, {Dexter}, {Dillon}, {Ely}, {Ewall-Wice}, {Fagnoni}, {Fialkov}, {Fritz}, {Furlanetto}, {Gale-Sides}, {Garsden}, {Glendenning}, {Gorce}, {Gorthi}, {Greig}, {Grobbelaar}, {Halday}, {Hazelton}, {Heimersheim}, {Hewitt}, {Hickish}, {Jacobs}, {Julius}, {Kern}, {Kerrigan}, {Kittiwisit}, {Kohn}, {Kolopanis}, {Lanman}, {La Plante}, {Lewis}, {Liu}, {Loots}, {Ma}, {MacMahon}, {Malan}, {Malgas}, {Malgas}, {Maree}, {Marero}, {Martinot}, {McBride}, {Mesinger}, {Mirocha}, {Molewa}, {Morales}, {Mosiane}, {Mu{\~n}oz}, {Murray}, {Nagpal}, {Neben}, {Nikolic}, {Nunhokee}, {Nuwegeld}, {Parsons}, {Pascua}, {Patra}, {Pieterse}, {Qin}, {Razavi-Ghods}, {Robnett}, {Rosie}, {Santos}, {Sims}, {Singh}, {Smith}, {Swarts}, {Tan},
  {Thyagarajan}, {Wilensky}, {Williams}, {van Wyngaarden}, \& {Zheng}}]{2023:hera}
{HERA Collaboration}, {Abdurashidova}, Z., {Adams}, T., {et~al.} 2023, \apj, 945, 124

\bibitem[{{Hinshaw} {et~al.}(2013){Hinshaw}, {Larson}, {Komatsu}, {Spergel}, {Bennett}, {Dunkley}, {Nolta}, {Halpern}, {Hill}, {Odegard}, {Page}, {Smith}, {Weiland}, {Gold}, {Jarosik}, {Kogut}, {Limon}, {Meyer}, {Tucker}, {Wollack}, \& {Wright}}]{2013:hinshaw}
{Hinshaw}, G., {Larson}, D., {Komatsu}, E., {et~al.} 2013, \apjs, 208, 19

\bibitem[{Jelić {et~al.}(2010)Jelić, Zaroubi, Labropoulos, Bernardi, de~Bruyn, \& Koopmans}]{2010:jelic}
Jelić, V., Zaroubi, S., Labropoulos, P., {et~al.} 2010, \mnras, 409, 1647

\bibitem[{Jordan {et~al.}(2017)Jordan, Murray, Trott, Wayth, Mitchell, Rahimi, Pindor, Procopio, \& Morgan}]{2017:jordan}
Jordan, C.~H., Murray, S., Trott, C.~M., {et~al.} 2017, \mnras, 471, 3974

\bibitem[{{Kern} {et~al.}(2020){Kern}, {Parsons}, {Dillon}, {Lanman}, {Liu}, {Bull}, {Ewall-Wice}, {Abdurashidova}, {Aguirre}, {Alexander}, {Ali}, {Balfour}, {Beardsley}, {Bernardi}, {Bowman}, {Bradley}, {Burba}, {Carilli}, {Cheng}, {DeBoer}, {Dexter}, {de Lera Acedo}, {Fagnoni}, {Fritz}, {Furlanetto}, {Glendenning}, {Gorthi}, {Greig}, {Grobbelaar}, {Halday}, {Hazelton}, {Hewitt}, {Hickish}, {Jacobs}, {Julius}, {Kerrigan}, {Kittiwisit}, {Kohn}, {Kolopanis}, {La Plante}, {Lekalake}, {MacMahon}, {Malan}, {Malgas}, {Maree}, {Martinot}, {Matsetela}, {Mesinger}, {Molewa}, {Morales}, {Mosiane}, {Murray}, {Neben}, {Parsons}, {Patra}, {Pieterse}, {Pober}, {Razavi-Ghods}, {Ringuette}, {Robnett}, {Rosie}, {Sims}, {Smith}, {Syce}, {Thyagarajan}, {Williams}, \& {Zheng}}]{2020:Kern}
{Kern}, N.~S., {Parsons}, A.~R., {Dillon}, J.~S., {et~al.} 2020, \apj, 888, 70

\bibitem[{{Koopmans}(2010)}]{2010:koopmans}
{Koopmans}, L.~V.~E. 2010, \apj, 718, 963

\bibitem[{{Mertens} {et~al.}(2024){Mertens}, {Bobin}, \& {Carucci}}]{2024:mertens}
{Mertens}, F.~G., {Bobin}, J., \& {Carucci}, I.~P. 2024, \mnras, 527, 3517

\bibitem[{{Mertens} {et~al.}(2018){Mertens}, {Ghosh}, \& {Koopmans}}]{2018:mertens}
{Mertens}, F.~G., {Ghosh}, A., \& {Koopmans}, L.~V.~E. 2018, \mnras, 478, 3640

\bibitem[{{Mertens} {et~al.}(2020){Mertens}, {Mevius}, {Koopmans}, {Offringa}, {Mellema}, {Zaroubi}, {Brentjens}, {Gan}, {Gehlot}, {Pandey}, {Sardarabadi}, {Vedantham}, {Yatawatta}, {Asad}, {Ciardi}, {Chapman}, {Gazagnes}, {Ghara}, {Ghosh}, {Giri}, {Iliev}, {Jeli{\'c}}, {Kooistra}, {Mondal}, {Schaye}, \& {Silva}}]{2020:mertens}
{Mertens}, F.~G., {Mevius}, M., {Koopmans}, L.~V.~E., {et~al.} 2020, \mnras, 493, 1662

\bibitem[{{Mertens} {et~al.}(2025){Mertens}, {Mevius}, {Koopmans}, {Offringa}, {Zaroubi}, {Acharya}, {Brackenhoff}, {Ceccotti}, {Chapman}, {Chege}, {Ciardi}, {Ghara}, {Ghosh}, {Giri}, {Hothi}, {H{\"o}fer}, {Iliev}, {Jeli{\'c}}, {Ma}, {Mellema}, {Munshi}, {Pandey}, \& {Yatawatta}}]{2025:mertens}
{Mertens}, F.~G., {Mevius}, M., {Koopmans}, L.~V.~E., {et~al.} 2025, \aap, 698, A186

\bibitem[{{Mevius} {et~al.}(2022){Mevius}, {Mertens}, {Koopmans}, {Offringa}, {Yatawatta}, {Brentjens}, {Chapman}, {Ciardi}, {Gan}, {Gehlot}, {Ghara}, {Ghosh}, {Giri}, {Iliev}, {Mellema}, {Pandey}, \& {Zaroubi}}]{2022:mevius}
{Mevius}, M., {Mertens}, F., {Koopmans}, L.~V.~E., {et~al.} 2022, \mnras, 509, 3693

\bibitem[{{Morales} {et~al.}(2012){Morales}, {Hazelton}, {Sullivan}, \& {Beardsley}}]{2012:morales}
{Morales}, M.~F., {Hazelton}, B., {Sullivan}, I., \& {Beardsley}, A. 2012, \apj, 752, 137

\bibitem[{{Morales} \& {Hewitt}(2004)}]{2004:morales}
{Morales}, M.~F. \& {Hewitt}, J. 2004, \apj, 615, 7

\bibitem[{{Mouri Sardarabadi} \& {Koopmans}(2019)}]{2019:Sardarabadi}
{Mouri Sardarabadi}, A. \& {Koopmans}, L.~V.~E. 2019, \mnras, 483, 5480

\bibitem[{{Offringa} {et~al.}(2014){Offringa}, {McKinley}, {Hurley-Walker}, {Briggs}, {Wayth}, {Kaplan}, {Bell}, {Feng}, {Neben}, {Hughes}, {Rhee}, {Murphy}, {Bhat}, {Bernardi}, {Bowman}, {Cappallo}, {Corey}, {Deshpande}, {Emrich}, {Ewall-Wice}, {Gaensler}, {Goeke}, {Greenhill}, {Hazelton}, {Hindson}, {Johnston-Hollitt}, {Jacobs}, {Kasper}, {Kratzenberg}, {Lenc}, {Lonsdale}, {Lynch}, {McWhirter}, {Mitchell}, {Morales}, {Morgan}, {Kudryavtseva}, {Oberoi}, {Ord}, {Pindor}, {Procopio}, {Prabu}, {Riding}, {Roshi}, {Shankar}, {Srivani}, {Subrahmanyan}, {Tingay}, {Waterson}, {Webster}, {Whitney}, {Williams}, \& {Williams}}]{2014:offringa}
{Offringa}, A.~R., {McKinley}, B., {Hurley-Walker}, N., {et~al.} 2014, \mnras, 444, 606

\bibitem[{{Offringa} {et~al.}(2019{\natexlab{a}}){Offringa}, {Mertens}, \& {Koopmans}}]{2019a:offringa}
{Offringa}, A.~R., {Mertens}, F., \& {Koopmans}, L.~V.~E. 2019{\natexlab{a}}, \mnras, 484, 2866

\bibitem[{{Offringa} {et~al.}(2019{\natexlab{b}}){Offringa}, {Mertens}, {van der Tol}, {Veenboer}, {Gehlot}, {Koopmans}, \& {Mevius}}]{2019b:offringa}
{Offringa}, A.~R., {Mertens}, F., {van der Tol}, S., {et~al.} 2019{\natexlab{b}}, \aap, 631, A12

\bibitem[{{Offringa} {et~al.}(2012){Offringa}, {van de Gronde}, \& {Roerdink}}]{2012:offringa}
{Offringa}, A.~R., {van de Gronde}, J.~J., \& {Roerdink}, J.~B.~T.~M. 2012, \aap, 539, A95

\bibitem[{{Ouchi} {et~al.}(2010){Ouchi}, {Shimasaku}, {Furusawa}, {Saito}, {Yoshida}, {Akiyama}, {Ono}, {Yamada}, {Ota}, {Kashikawa}, {Iye}, {Kodama}, {Okamura}, {Simpson}, \& {Yoshida}}]{2010:Ouchi}
{Ouchi}, M., {Shimasaku}, K., {Furusawa}, H., {et~al.} 2010, \apj, 723, 869

\bibitem[{{Patil} {et~al.}(2017){Patil}, {Yatawatta}, {Koopmans}, {de Bruyn}, {Brentjens}, {Zaroubi}, {Asad}, {Hatef}, {Jeli{\'c}}, {Mevius}, {Offringa}, {Pandey}, {Vedantham}, {Abdalla}, {Brouw}, {Chapman}, {Ciardi}, {Gehlot}, {Ghosh}, {Harker}, {Iliev}, {Kakiichi}, {Majumdar}, {Mellema}, {Silva}, {Schaye}, {Vrbanec}, \& {Wijnholds}}]{2017:patil}
{Patil}, A.~H., {Yatawatta}, S., {Koopmans}, L.~V.~E., {et~al.} 2017, \apj, 838, 65

\bibitem[{Patil {et~al.}(2016)Patil, Yatawatta, Zaroubi, Koopmans, de~Bruyn, Jelić, Ciardi, Iliev, Mevius, Pandey, \& Gehlot}]{2016:patil}
Patil, A.~H., Yatawatta, S., Zaroubi, S., {et~al.} 2016, \mnras, 463, 4317

\bibitem[{{Planck Collaboration} {et~al.}(2020){Planck Collaboration}, {Aghanim}, {Akrami}, {Ashdown}, {Aumont}, {Baccigalupi}, {Ballardini}, {Banday}, {Barreiro}, {Bartolo}, {Basak}, {Battye}, {Benabed}, {Bernard}, {Bersanelli}, {Bielewicz}, {Bock}, {Bond}, {Borrill}, {Bouchet}, {Boulanger}, {Bucher}, {Burigana}, {Butler}, {Calabrese}, {Cardoso}, {Carron}, {Challinor}, {Chiang}, {Chluba}, {Colombo}, {Combet}, {Contreras}, {Crill}, {Cuttaia}, {de Bernardis}, {de Zotti}, {Delabrouille}, {Delouis}, {Di Valentino}, {Diego}, {Dor{\'e}}, {Douspis}, {Ducout}, {Dupac}, {Dusini}, {Efstathiou}, {Elsner}, {En{\ss}lin}, {Eriksen}, {Fantaye}, {Farhang}, {Fergusson}, {Fernandez-Cobos}, {Finelli}, {Forastieri}, {Frailis}, {Fraisse}, {Franceschi}, {Frolov}, {Galeotta}, {Galli}, {Ganga}, {G{\'e}nova-Santos}, {Gerbino}, {Ghosh}, {Gonz{\'a}lez-Nuevo}, {G{\'o}rski}, {Gratton}, {Gruppuso}, {Gudmundsson}, {Hamann}, {Handley}, {Hansen}, {Herranz}, {Hildebrandt}, {Hivon}, {Huang}, {Jaffe}, {Jones}, {Karakci}, {Keih{\"a}nen},
  {Keskitalo}, {Kiiveri}, {Kim}, {Kisner}, {Knox}, {Krachmalnicoff}, {Kunz}, {Kurki-Suonio}, {Lagache}, {Lamarre}, {Lasenby}, {Lattanzi}, {Lawrence}, {Le Jeune}, {Lemos}, {Lesgourgues}, {Levrier}, {Lewis}, {Liguori}, {Lilje}, {Lilley}, {Lindholm}, {L{\'o}pez-Caniego}, {Lubin}, {Ma}, {Mac{\'\i}as-P{\'e}rez}, {Maggio}, {Maino}, {Mandolesi}, {Mangilli}, {Marcos-Caballero}, {Maris}, {Martin}, {Martinelli}, {Mart{\'\i}nez-Gonz{\'a}lez}, {Matarrese}, {Mauri}, {McEwen}, {Meinhold}, {Melchiorri}, {Mennella}, {Migliaccio}, {Millea}, {Mitra}, {Miville-Desch{\^e}nes}, {Molinari}, {Montier}, {Morgante}, {Moss}, {Natoli}, {N{\o}rgaard-Nielsen}, {Pagano}, {Paoletti}, {Partridge}, {Patanchon}, {Peiris}, {Perrotta}, {Pettorino}, {Piacentini}, {Polastri}, {Polenta}, {Puget}, {Rachen}, {Reinecke}, {Remazeilles}, {Renzi}, {Rocha}, {Rosset}, {Roudier}, {Rubi{\~n}o-Mart{\'\i}n}, {Ruiz-Granados}, {Salvati}, {Sandri}, {Savelainen}, {Scott}, {Shellard}, {Sirignano}, {Sirri}, {Spencer}, {Sunyaev}, {Suur-Uski}, {Tauber}, {Tavagnacco},
  {Tenti}, {Toffolatti}, {Tomasi}, {Trombetti}, {Valenziano}, {Valiviita}, {Van Tent}, {Vibert}, {Vielva}, {Villa}, {Vittorio}, {Wandelt}, {Wehus}, {White}, {White}, {Zacchei}, \& {Zonca}}]{2020:Planck}
{Planck Collaboration}, {Aghanim}, N., {Akrami}, Y., {et~al.} 2020, \aap, 641, A6

\bibitem[{{Rogers} \& {Bowman}(2008)}]{2008:rogers}
{Rogers}, A. E.~E. \& {Bowman}, J.~D. 2008, \aj, 136, 641

\bibitem[{{Santos} {et~al.}(2005){Santos}, {Cooray}, \& {Knox}}]{2005:santos}
{Santos}, M.~G., {Cooray}, A., \& {Knox}, L. 2005, \apj, 625, 575

\bibitem[{{Singh} {et~al.}(2018){Singh}, {Subrahmanyan}, {Udaya Shankar}, {Sathyanarayana Rao}, {Fialkov}, {Cohen}, {Barkana}, {Girish}, {Raghunathan}, {Somashekar}, \& {Srivani}}]{2018:singh}
{Singh}, S., {Subrahmanyan}, R., {Udaya Shankar}, N., {et~al.} 2018, \apj, 858, 54

\bibitem[{{Smirnov}(2011)}]{2011:smirnov}
{Smirnov}, O.~M. 2011, \aap, 527, A106

\bibitem[{Spinelli {et~al.}(2018)Spinelli, Bernardi, \& Santos}]{2018:spinelli}
Spinelli, M., Bernardi, G., \& Santos, M.~G. 2018, \mnras, 479, 275

\bibitem[{{Stark} {et~al.}(2010){Stark}, {Ellis}, {Chiu}, {Ouchi}, \& {Bunker}}]{2010:stark}
{Stark}, D.~P., {Ellis}, R.~S., {Chiu}, K., {Ouchi}, M., \& {Bunker}, A. 2010, \mnras, 408, 1628

\bibitem[{{Tingay} {et~al.}(2013){Tingay}, {Goeke}, {Bowman}, {Emrich}, {Ord}, {Mitchell}, {Morales}, {Booler}, {Crosse}, {Wayth}, {Lonsdale}, {Tremblay}, {Pallot}, {Colegate}, {Wicenec}, {Kudryavtseva}, {Arcus}, {Barnes}, {Bernardi}, {Briggs}, {Burns}, {Bunton}, {Cappallo}, {Corey}, {Deshpande}, {Desouza}, {Gaensler}, {Greenhill}, {Hall}, {Hazelton}, {Herne}, {Hewitt}, {Johnston-Hollitt}, {Kaplan}, {Kasper}, {Kincaid}, {Koenig}, {Kratzenberg}, {Lynch}, {Mckinley}, {Mcwhirter}, {Morgan}, {Oberoi}, {Pathikulangara}, {Prabu}, {Remillard}, {Rogers}, {Roshi}, {Salah}, {Sault}, {Udaya-Shankar}, {Schlagenhaufer}, {Srivani}, {Stevens}, {Subrahmanyan}, {Waterson}, {Webster}, {Whitney}, {Williams}, {Williams}, \& {Wyithe}}]{2013:tingay}
{Tingay}, S.~J., {Goeke}, R., {Bowman}, J.~D., {et~al.} 2013, \pasa, 30, e007

\bibitem[{{Trott} {et~al.}(2020){Trott}, {Jordan}, {Midgley}, {Barry}, {Greig}, {Pindor}, {Cook}, {Sleap}, {Tingay}, {Ung}, {Hancock}, {Williams}, {Bowman}, {Byrne}, {Chokshi}, {Hazelton}, {Hasegawa}, {Jacobs}, {Joseph}, {Li}, {Line}, {Lynch}, {McKinley}, {Mitchell}, {Morales}, {Ouchi}, {Pober}, {Rahimi}, {Takahashi}, {Wayth}, {Webster}, {Wilensky}, {Wyithe}, {Yoshiura}, {Zhang}, \& {Zheng}}]{2020:trott}
{Trott}, C.~M., {Jordan}, C.~H., {Midgley}, S., {et~al.} 2020, \mnras, 493, 4711

\bibitem[{{van der Tol} {et~al.}(2018){van der Tol}, {Veenboer}, \& {Offringa}}]{2018:vandertol}
{van der Tol}, S., {Veenboer}, B., \& {Offringa}, A.~R. 2018, \aap, 616, A27

\bibitem[{{van Diepen} {et~al.}(2018){van Diepen}, {Dijkema}, \& {Offringa}}]{2018:vandiepen}
{van Diepen}, G., {Dijkema}, T.~J., \& {Offringa}, A. 2018, {DPPP: Default Pre-Processing Pipeline}, Astrophysics Source Code Library, record ascl:1804.003

\bibitem[{{van Haarlem} {et~al.}(2013){van Haarlem}, {Wise}, {Gunst}, {Heald}, {McKean}, {Hessels}, {de Bruyn}, {Nijboer}, {Swinbank}, {Fallows}, {Brentjens}, {Nelles}, {Beck}, {Falcke}, {Fender}, {H{\"o}randel}, {Koopmans}, {Mann}, {Miley}, {R{\"o}ttgering}, {Stappers}, {Wijers}, {Zaroubi}, {van den Akker}, {Alexov}, {Anderson}, {Anderson}, {van Ardenne}, {Arts}, {Asgekar}, {Avruch}, {Batejat}, {B{\"a}hren}, {Bell}, {Bell}, {van Bemmel}, {Bennema}, {Bentum}, {Bernardi}, {Best}, {B{\^\i}rzan}, {Bonafede}, {Boonstra}, {Braun}, {Bregman}, {Breitling}, {van de Brink}, {Broderick}, {Broekema}, {Brouw}, {Br{\"u}ggen}, {Butcher}, {van Cappellen}, {Ciardi}, {Coenen}, {Conway}, {Coolen}, {Corstanje}, {Damstra}, {Davies}, {Deller}, {Dettmar}, {van Diepen}, {Dijkstra}, {Donker}, {Doorduin}, {Dromer}, {Drost}, {van Duin}, {Eisl{\"o}ffel}, {van Enst}, {Ferrari}, {Frieswijk}, {Gankema}, {Garrett}, {de Gasperin}, {Gerbers}, {de Geus}, {Grie{\ss}meier}, {Grit}, {Gruppen}, {Hamaker}, {Hassall}, {Hoeft}, {Holties},
  {Horneffer}, {van der Horst}, {van Houwelingen}, {Huijgen}, {Iacobelli}, {Intema}, {Jackson}, {Jelic}, {de Jong}, {Juette}, {Kant}, {Karastergiou}, {Koers}, {Kollen}, {Kondratiev}, {Kooistra}, {Koopman}, {Koster}, {Kuniyoshi}, {Kramer}, {Kuper}, {Lambropoulos}, {Law}, {van Leeuwen}, {Lemaitre}, {Loose}, {Maat}, {Macario}, {Markoff}, {Masters}, {McFadden}, {McKay-Bukowski}, {Meijering}, {Meulman}, {Mevius}, {Middelberg}, {Millenaar}, {Miller-Jones}, {Mohan}, {Mol}, {Morawietz}, {Morganti}, {Mulcahy}, {Mulder}, {Munk}, {Nieuwenhuis}, {van Nieuwpoort}, {Noordam}, {Norden}, {Noutsos}, {Offringa}, {Olofsson}, {Omar}, {Orr{\'u}}, {Overeem}, {Paas}, {Pandey-Pommier}, {Pandey}, {Pizzo}, {Polatidis}, {Rafferty}, {Rawlings}, {Reich}, {de Reijer}, {Reitsma}, {Renting}, {Riemers}, {Rol}, {Romein}, {Roosjen}, {Ruiter}, {Scaife}, {van der Schaaf}, {Scheers}, {Schellart}, {Schoenmakers}, {Schoonderbeek}, {Serylak}, {Shulevski}, {Sluman}, {Smirnov}, {Sobey}, {Spreeuw}, {Steinmetz}, {Sterks}, {Stiepel}, {Stuurwold},
  {Tagger}, {Tang}, {Tasse}, {Thomas}, {Thoudam}, {Toribio}, {van der Tol}, {Usov}, {van Veelen}, {van der Veen}, {ter Veen}, {Verbiest}, {Vermeulen}, {Vermaas}, {Vocks}, {Vogt}, {de Vos}, {van der Wal}, {van Weeren}, {Weggemans}, {Weltevrede}, {White}, {Wijnholds}, {Wilhelmsson}, {Wucknitz}, {Yatawatta}, {Zarka}, {Zensus}, \& {van Zwieten}}]{2013:van_Haarlem}
{van Haarlem}, M.~P., {Wise}, M.~W., {Gunst}, A.~W., {et~al.} 2013, \aap, 556, A2

\bibitem[{Vedantham \& Koopmans(2016)}]{2016:vedantham}
Vedantham, H.~K. \& Koopmans, L. V.~E. 2016, \mnras, 458, 3099

\bibitem[{Veenboer(2021)}]{2021:veenboer}
Veenboer, A. 2021, Phd-thesis - research and graduation internal, Vrije Universiteit Amsterdam

\bibitem[{{Wijnholds} {et~al.}(2016){Wijnholds}, {Grobler}, \& {Smirnov}}]{2016:Wijnholds}
{Wijnholds}, S.~J., {Grobler}, T.~L., \& {Smirnov}, O.~M. 2016, \mnras, 457, 2331

\bibitem[{{Wilensky} {et~al.}(2019){Wilensky}, {Morales}, {Hazelton}, {Barry}, {Byrne}, \& {Roy}}]{2019:wilensky}
{Wilensky}, M.~J., {Morales}, M.~F., {Hazelton}, B.~J., {et~al.} 2019, \pasp, 131, 114507

\bibitem[{Yatawatta(2015)}]{2015:yatawatta}
Yatawatta, S. 2015, \mnras, 449, 4506

\bibitem[{{Yatawatta}(2016)}]{2016:yatawatta}
{Yatawatta}, S. 2016, arXiv e-prints, arXiv:1605.09219

\bibitem[{Zarka {et~al.}(2020)Zarka, Denis, Tagger, Girard, Coffre, Dumez-Viou, Taffoureau, Charrier, Bondonneau, Briand, Casoli, Cecconi, Cognard, Corbel, Dallier, Ferrari, Grie{\ss}meier, Loh, Martin, Pommier, Benoit, Tasse, Theureau, Tremou, Capayrou, Censier, Cottet, Etieve, Floquet, Garnier, Georges, Gond, Jacquet, Joly, Gall, Roziere, Thetas, Vimon, Guilbeau, Sandre, Courte, Desvignes, Marchand, Alves, Azarian, Bonnassieux, Brionne, Chassande-Mottin, Combes, Cornilleau-Wehrlin, Decoene, G{\'e}rard, Gusdorf, Lamy, Masson, Mertens, Mottez, Petri, Vilmer, Weber, Hellbourg, Konovalenko, Koopmans, Tokarsky, Ulyanov, Vermeulen, \& Zakharenko}]{2020:zarka}
Zarka, P., Denis, L., Tagger, M., {et~al.} 2020, in {URSI GASS 2020}, Rome, Italy

\bibitem[{{Zheng} {et~al.}(2017){Zheng}, {Tegmark}, {Dillon}, {Kim}, {Liu}, {Neben}, {Jonas}, {Reich}, \& {Reich}}]{2017:zheng}
{Zheng}, H., {Tegmark}, M., {Dillon}, J.~S., {et~al.} 2017, \mnras, 464, 3486

\end{thebibliography}

\bibliographystyle{aa}

\begin{appendix}

\section{Impact of diffuse Galactic emission on DI-Gain calibration in image space}
\label{sec:appendix_c}
\begin{figure*}[h!]
    \centering
    \includegraphics[trim={1cm 0cm 1cm 5cm},clip, scale=0.3]{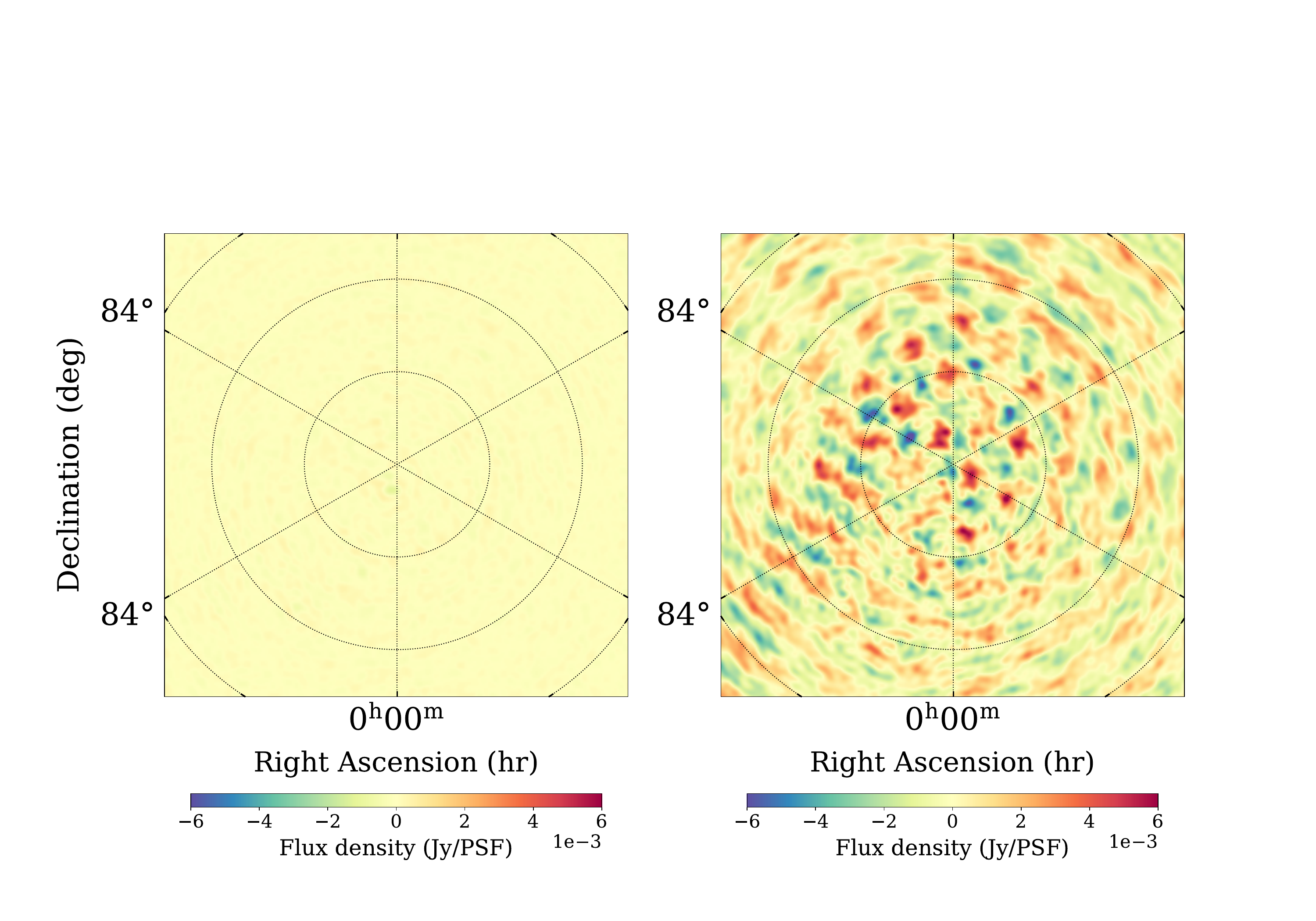}
    \caption{The left panel shows the residual emission obtained by differencing the DI-gain calibrated sky from the true sky in our ground truth scenario (i.e., when calibrating with a matched PSO sky model). The right panel shows the residual emission obtained when calibrating with a mismatched sky model.}    
    \label{fig:diff_img_ps}
\end{figure*}

We generated images of the true sky, which consists of point sources and thermal noise in our ground truth scenario and point sources, diffuse Galactic emission, and thermal noise in the other case. We then created DI-gain calibrated images for both scenarios and subtracted them from their respective true sky images. The Figure \ref{fig:diff_img_ps} shows the resulting residuals: the left panel corresponds to calibration with a matched PSO sky model, while the right panel corresponds to DI-gain calibration with a mismatched sky model, where diffuse Galactic emission was missing during the calibration process. This highlights how diffuse Galactic emission is absorbed into the DI-gain solutions and appears as residual emission in image space.

\section{Systematic bias in stacked power spectra}
\label{sec:appendix_a}
We combine three simulated observations taken at different local sidereal times separated by 4 hours, by using inverse variance weighting to optimally average the data in visibility space. The systematic bias as a fraction of thermal noise is shown for stacks of 1, 2 and 3 simulated observations in Figure \ref{fig:stack_complete} when DI-gain calibrating visibilities with a matched PSO sky model and in Figure \ref{fig:stack_incomplete} when DI-gain calibrating with a mismatched sky model. The systematic bias in the foreground region for $k_{\parallel}< 0.2 \, h\mathrm{MPc}^{-1}$ is at least one magnitude higher when DI-gain calibrating with a mismatched sky model, which is missing the diffuse Galactic emission. As will be shown in the next section \ref{sec:appendix_b}, the bias in the foreground region is highly coherent, therefore our foreground removal method, Gaussian Process Regression (GPR), has proven to be able to remove this excess power.

\begin{figure*}[h!]
    \centering
    \includegraphics[trim={1cm 0 1cm 0},clip, scale=0.4]{202507_psstack_complete_skymodel_seismic.png}
    \caption{The systematic bias as a fraction of the thermal noise for matched PSO sky model DI-gain calibration, when the cylindrical Stokes I power spectra are coherently averaged for 1, 2 and 3 observations, respectively. Each observation is 4 hours long and separated by 4 hours in local sidereal time, in order to cover a full day of observation for a symmetric array.}
    \label{fig:stack_complete}
\end{figure*}

\begin{figure*}[h!]
    \centering
    \includegraphics[trim={1cm 0 1cm 0},clip, scale=0.4]{202507_psstack_incomplete_skymodel_seismic.png}
    \caption{The systematic bias as a fraction of the thermal noise for mismatched sky model DI-gain calibration, when the cylindrical Stokes I power spectra are coherently averaged for 1, 2 and 3 days, respectively. Each observation is 4 hours long and separated by 4 hours in local sidereal time, in order to cover a full day of observation for a symmetric array.}    
    \label{fig:stack_incomplete}
\end{figure*}

\section{Average cross-coherence}
\label{sec:appendix_b}
The average cross-coherence is calculated by computing the cross-coherence, as defined by Equation \ref{eq:coherence}, for all pairs of observations and then averaging the results. The left panel of figure \ref{fig:avg_cross_coh} shows the average cross-coherence of 2D power spectra derived from visibilities DI-gain calibrated using a matched PSO sky model. The middle panel presents the same analysis for visibilities DI-gain calibrated with an mismatched sky model, which omits the diffuse Galactic emission component. The power spectrum bins in the lowest $k_{\parallel}$-bin show high coherence due to the presence of foregrounds. Additionally, regions at low $k_{\perp}$, which correspond to power spectrum bins associated with $uv$-cells containing fewer visibilities, also show increased coherence. At first glance, there appears to be no significant difference between the two. However, a comparison of the average cross-coherence between the middle panel (mismatched sky model DI-gain calibration) and the left panel (matched PSO sky model DI-gain calibration) reveals that certain power spectrum bins associated with short baselines exhibit higher coherence when calibrated with the mismatched sky model. This suggests the presence of coherent excess noise that might not average down in the same manner as thermal noise.
\begin{figure*}[h!]
    \centering
    \includegraphics[trim={1cm 0 1cm 0},clip, scale=0.35]{202507_difference_coherence_incomplete_skymodel_to_complete_skymodel.png}
    \caption{The average cross-coherence between three different observations covering different local sidereal time (LST) ranges for power spectra obtained from visibilities calibrated with a matched PSO sky model in the left panel, calibrated with a mismatched sky model, that is missing the diffuse Galactic emission component, in the middle panel and the difference between them in the right panel. There is an increase in coherence in $(k_{\perp}, k_{\parallel})$-bins corresponding to short baselines $|\mathbf{u}| < 100$, showing that DI-gain calibration errors due to missing Galactic diffuse emission are partly coherent.}
    \label{fig:avg_cross_coh}
\end{figure*}

\end{appendix}
\end{document}